\documentclass[10pt,twocolumn,letterpaper]{article}

\usepackage{iccv}              
\usepackage{amsmath,amssymb,amsfonts}
\usepackage{algorithmic}
\usepackage{graphicx}
\usepackage{textcomp}
\usepackage{xcolor}
\usepackage{booktabs}
\usepackage{multirow}
\usepackage{array}
\usepackage{float}
\usepackage{breakcites}

\definecolor{iccvblue}{rgb}{0.21,0.49,0.74}
\usepackage[pagebackref,breaklinks,colorlinks,allcolors=iccvblue]{hyperref}

\def\paperID{11534} 
\def\confName{ICCV}
\def\confYear{2025}

\title{HyTIP: Hybrid Temporal Information Propagation \\for Masked Conditional Residual Video Coding}

\author{
Yi-Hsin Chen\textsuperscript{1}     \qquad
Yi-Chen Yao\textsuperscript{1}      \qquad
Kuan-Wei Ho\textsuperscript{1}      \qquad
Chun-Hung Wu\textsuperscript{1}     \qquad
Huu-Tai Phung\textsuperscript{1} \\ \qquad
Martin Benjak\textsuperscript{2}    \qquad
Jörn Ostermann\textsuperscript{2}   \qquad
Wen-Hsiao Peng\textsuperscript{1} \\ \\
\textsuperscript{1} National Yang Ming Chiao Tung University, Taiwan \\
\textsuperscript{2} Leibniz Universität Hannover, Germany
}

\newcommand{\new}[1]{\textcolor{black}{#1}}

\usepackage[capitalize]{cleveref}
\crefname{section}{Sec.}{Secs.}
\Crefname{section}{Section}{Sections}
\Crefname{table}{Table}{Tables}
\crefname{table}{Tab.}{Tabs.}

\newcommand{\beginsupplement}{%
        \setcounter{table}{0}
        \renewcommand{\thetable}{A\arabic{table}}%
        \setcounter{figure}{0}
        \renewcommand{\thefigure}{A\arabic{figure}}%
        \setcounter{section}{0}
        \renewcommand{\thesection}{A\arabic{section}}%

}

\begin{document}
\maketitle

\begin{abstract}
Most frame-based learned video codecs can be interpreted as recurrent neural networks (RNNs) propagating reference information along the temporal dimension. This work revisits the limitations of the current approaches from an RNN perspective. The output-recurrence methods, which propagate decoded frames, are intuitive but impose dual constraints on the output decoded frames, leading to suboptimal rate-distortion performance. In contrast, the hidden-to-hidden connection approaches, which propagate latent features within the RNN, offer greater flexibility but require large buffer sizes. To address these issues, we propose HyTIP, a learned video coding framework that combines both mechanisms. Our hybrid buffering strategy uses explicit decoded frames and a small number of implicit latent features to achieve competitive coding performance. Experimental results show that our HyTIP outperforms the sole use of either output-recurrence or hidden-to-hidden approaches. Furthermore, it achieves comparable performance to state-of-the-art methods but with a much smaller buffer size, and outperforms VTM 17.0 (Low-delay B) in terms of PSNR-RGB and MS-SSIM-RGB. \new{The source code of HyTIP is available at \href{https://github.com/NYCU-MAPL/HyTIP}{https://github.com/NYCU-MAPL/HyTIP}.}

\end{abstract}  
\section{Introduction}
\label{sec:intro}
Most modern learned video coding schemes bear an interpretation of implementing a recurrent neural network (RNN) along the temporal dimension. The encoding of an input video is usually performed frame-by-frame, with a shared model, composed of an encoder and a decoder, employed at each time step to convert an input video frame into its frame latents as the encoder output and to reconstruct the input frame approximately as the decoder output. The process involves leveraging the past information in the form of decoded frames and/or their latent features to encode the resulting frame latents in a rate-distortion-optimized manner. As with an RNN, the frame-by-frame coding and model weight sharing across time steps are essential for coding variable-length video sequences. 


\begin{figure}[t!]
    \centering
    \vspace{-0.3cm}
    \subfloat[Output-recurrence connection with explicit buffer]{
        \includegraphics[width=0.94\linewidth, trim=0 192 0 0, clip]{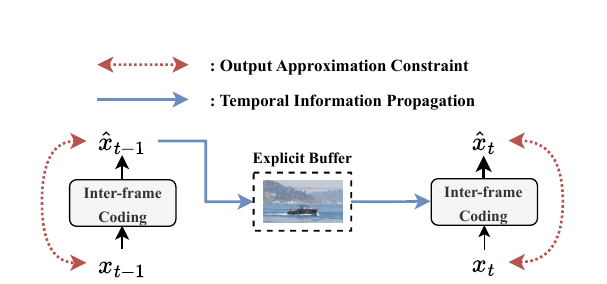}
    }
    \hfill 
    \subfloat[Hidden-to-hidden connection with implicit buffer]{
        \includegraphics[width=0.94\linewidth, trim=0 112 0 135, clip]{Figure/architecture/teaser_bosphorus.pdf}
    }
    \hfill 
    \subfloat[Our proposed hybrid scheme with hybrid buffer]{
        \includegraphics[width=0.94\linewidth, trim=0 30 0 220, clip]{Figure/architecture/teaser_bosphorus.pdf}
    }
    \caption{Comparison of different temporal information propagation mechanisms in learned video compression.}
    \vspace{-0.4cm}
    \label{fig:teaser}
\end{figure}

Based on how temporal information is propagated through the RNN, these learned video codecs can be broadly categorized into two major approaches that feature either the output-recurrence~\cite{dvc, epa, rafc, mlvc, nvc, ssf, fvc, modenet, dcvc, canfvc, lvcyuv, pcs22, c2f, crc, maskcrt, mmsp24} or the hidden-to-hidden connection~\cite{tcm, dc, fm, sdd, pqa}. As depicted in Fig.~\ref{fig:teaser}~(a), the former approach, which adopts the output-recurrence connection, follows the same coding architecture as traditional video codecs~\cite{avc, hevc, vvc}, where the previously decoded frame is \textit{explicitly} buffered as a reference frame to encode the next frame. However, RNNs with only the output-recurrence connection face the challenge that the model output must closely resemble the ground truth, i.e. the input frame in the present context, while being a sufficient summary of the useful information from the past. This dual requirement leads to a compromise between preserving the past temporal information sufficiently and approximating the input frame accurately, thereby suboptimal rate-distortion performance.

To address the limitations of the output-recurrence design, the second approach aggregates temporal information by a hidden-to-hidden connection~\cite{tcm, dc, fm, sdd, pqa, mimt, vct}. Instead of storing the previously decoded frame, it propagates the latent features in the RNN across time steps (see Fig.~\ref{fig:teaser} (b)). In particular, the extraction and propagation of latent features are completely learned. As a result, there is little control over what information is stored, rendering the buffer's content non-explainable. To distinguish between these two buffering strategies, we refer to the buffer that stores latent features as the \textit{implicit buffer}, in contrast to the \textit{explicit buffer} that stores the previously decoded frame for the output-recurrence connection. Note that these temporally propagated features are not specifically constrained to approximate any input frame, although the output frame decoded from these features must closely resemble the input frame. In other words, these latent features may buffer more information from the past than is needed to reconstruct the current input frame. This design allows for more flexible temporal modeling, potentially leading to better coding efficiency. Many state-of-the-art neural video codecs belong to this category. Although some of them additionally buffer the previously decoded frame for motion estimation, the buffered frame is not directly involved as contextual information for inter-frame coding. 

\textcolor{black}{Methods adopting an implicit buffering strategy usually come at the cost of substantial memory requirements.}
For example, some~\cite{tcm, dc, fm, sdd, pqa} require the storage of at least 48 high-resolution feature maps that are of the same spatial resolution as the input frame. Generally, storing more contextual information from the past helps to improve coding efficiency. But, the data-driven nature often results in non-compact features being stored.


Recognizing the issues inherent in solely using either the output-recurrence or hidden-to-hidden connection, we propose a learned video compression framework termed HyTIP. It combines both the output-recurrence and hidden-to-hidden connections to harness the advantages of both approaches. Theoretically, RNNs that integrate both connection types are the most powerful in terms of expressivity~\cite{goodfellow2016deep}. As depicted in Fig.~\ref{fig:teaser} (c), we propagate the previously decoded frame as \textit{explicit} information along with only a small number of \textit{implicit} latent features. Our hybrid buffering strategy leverages the prior knowledge that the previously decoded frame typically has the highest correlation with the current input frame 
to reduce the reliance on implicit features. By doing so, we store only a few latent features to complement the previously decoded frame. Compared to the purely implicit buffering scheme, our hybrid approach significantly reduces the buffer size while achieving similar coding performance. Moreover, compared to the explicit buffering scheme, it is able to propagate the past information that is complementary to the previously decoded frame. Our hybrid buffering strategy is validated in a masked conditional residual coding framework~\cite{maskcrt}. 


In summary, our main contributions are as follows:
\begin{itemize}[]
\setlength\itemsep{.3em}
\item \textcolor{black}{To the best of our knowledge, this work is an early attempt to revisit the temporal information propagation mechanisms in learned video compression through an RNN perspective, analyzing their performance and buffer size trade-offs under a unified base codec.}
\item \textcolor{black}{This work presents the first masked conditional residual coding framework that adopts a hybrid buffering strategy. This strategy propagates both implicit and explicit temporal information for motion and inter-frame coding.}
\item Experimental results demonstrate that our HyTIP achieves comparable coding performance to the state-of-the-art methods, but requires only 14\% of their buffer size. Additionally, it outperforms VTM (Low-delay B) in terms of PSNR-RGB and MS-SSIM-RGB. 
\item Our simple yet efficient hybrid buffering scheme is easily extendable to other learned video codecs.
\end{itemize}

\section{Related Work}
\label{sec:related}
\subsection{Frame-based Learned Video Coding as RNN}

\begin{table}[t]
\fontsize{9pt}{9pt}\selectfont
\caption{Taxonomy of learned video compression works based on RNN type and the nature of buffered temporal information.}
\vspace{-0.2cm}
\centering
\setlength{\tabcolsep}{2pt}
\begin{tabular}{c|c|c}
RNN Type            &    \begin{tabular}[c]{@{}c@{}}Buffer for \\ Inter-frame Coding\end{tabular}    & Publications \\
\midrule
Output-recurrence       &      Explicit                                                                  &  \begin{tabular}[c]{@{}c@{}}\cite{rafc, ssf, fvc, modenet, dcvc, canfvc, lvcyuv, pcs22, c2f, crc, maskcrt, mmsp24} \\ \cite{mlvc, nvc, dvc, epa} \\ \cite{lvc, alphavc, elvc, lstvc} \end{tabular}            \\
\midrule
Hidden-to-hidden        & Implicit                                                                       &  \cite{tcm, dc, fm, sdd, pqa, mimt, vct}            \\
\midrule
\begin{tabular}[c]{@{}c@{}}Output-recurrence \\ + Hidden-to-hidden\end{tabular}   & Hybrid &  \cite{feedback, rlvc}           
\label{tab:related}
\end{tabular}
\vspace{-0.4cm}
\end{table}


Most learned video codecs that encode an input video frame-by-frame with an IPPP prediction structure can be thought of as a form of RNN. In RNN terminology, they propagate temporal information through either the output-recurrence or hidden-to-hidden connection. The former utilizes an \textit{explicit} buffer to store and propagate the decoded signals from the past, such as decoded frames and/or flow maps, while the latter employs an \textit{implicit} buffer for maintaining learned latent features. Table~\ref{tab:related} presents a taxonomy of the modern learned video codecs according to how they buffer and propagate temporal information. 

\noindent \textbf{Explicit Buffering Strategies:} Under the IPPP prediction structure, many learned video codecs~\cite{dvc, epa, rafc, mlvc, nvc, ssf, fvc, modenet, dcvc, canfvc, lvcyuv, pcs22, c2f, crc, maskcrt, mmsp24, elvc, alphavc, lvc, lstvc} \textit{explicitly} buffer the previously decoded frame(s) for inter-frame coding. 
Most approaches~\cite{dvc, epa, rafc, nvc, ssf, modenet, dcvc, lvcyuv, pcs22, crc, maskcrt, mmsp24, elvc, alphavc, lvc, lstvc} keeps only one previously decoded frame for motion estimation and inter-frame prediction because it usually correlates most strongly with the current input frame. However, some methods~\cite{mlvc} buffer multiple decoded frames and flow maps to construct a better temporal predictor. Similar predictive coding is also found in motion coding. For example, \cite{mlvc, canfvc, c2f} implement predictive motion coding based on decoded frames or flow maps. \textcolor{black}{While \cite{nvc, aaai20, elvc, alphavc, lvc, lstvc} propagate \textit{implicit} features as a temporal prior for probability distribution modeling in entropy coding, these features are not directly used to generate the decoded frame. Therefore, we consider these approaches to primarily propagate contextual information in a output-recurrence manner.} As discussed previously and from the perspective of the RNN structure, the dual role required of the decoded frame (or flow map) in the output-recurrence paradigm leads to suboptimal rate-distortion performance.

\noindent \textbf{Implicit Buffering Strategies:} In contrast to buffering decoded frames for inter-frame coding, \cite{tcm} buffers high-resolution, 64-channel latent features extracted from the inter-frame codec as contextual information for coding the next frame. In their work, when the number of feature maps stored is reduced from 64 to 9, a 2.5\% BD-rate drop is observed~\cite{tcm}. Most follow-up works thus maintained large buffer sizes. For example, \cite{dc, fm} use 48+ channels. Along this line of research, some recent studies~\cite{sdd, pqa} further increase the buffer size by introducing additional ConvLSTM layers~\cite{convlstm} to propagate long-term, high-resolution temporal information, in addition to the existing 64-channel latent features. In \cite{dc, fm}, the implicit buffering strategy is applied to both inter-frame and motion coding. While these methods achieve state-of-the-art coding results, their reliance on implicitly learned temporal information leads to the storage of non-compact latent features.

\noindent \textbf{Hybrid Buffering Strategies:}
There have also been attempts~\cite{feedback, rlvc} to combine the output-recurrence and hidden-to-hidden connections with a hybrid buffering strategy. All these works \textit{explicitly} buffer the previously decoded frame to construct a temporal predictor for inter-frame coding. Additionally, \cite{feedback} introduces a Conv-GRU~\cite{convgru} in the decoder to learn and propagate latent features for motion and inter-frame coding, while \cite{rlvc} use ConvLSTM layers~\cite{convlstm} in both the encoder and the decoder by the same token. 
\textcolor{black}{In comparison to these works, our HyTIP adopts a hybrid buffering strategy for both motion and inter-frame coding in a conditional residual coding framework. Notably, this work presents an in-depth study to shed light on the interactions between implicit and explicit buffers and their rate-distortion-complexity trade-offs.}

\subsection{Learned Video Coding Frameworks}
There are three mainstream approaches to incorporating temporal information for inter-frame coding: (1) residual coding, (2) conditional coding, and (3) conditional residual coding. Early learned video codecs~\cite{dvc, epa, rafc, mlvc, nvc, ssf, fvc, feedback, rlvc, lvc, aaai20} predominantly follow the notion of residual coding. Similar to traditional codecs, it encodes the frame difference $x_t-x_c$ between the input frame $x_t$ and its temporal predictor $x_c$ generated from the propagated temporal information. Rather than employing $x_c$ linearly, conditional coding~\cite{modenet, dcvc, canfvc, lvcyuv, elvc, alphavc, c2f, tcm,hem, mimt, vct, butterfly, mip, dc, fm, sdd, pqa} encodes $x_t$ by using $x_c$ as a condition signal for both the encoder and decoder, allowing $x_c$ to be utilized in a non-linear fashion for temporal prediction. This approach has been widely adopted by state-of-the-art learned video codecs. However, conditional coding could potentially suffer from the bottleneck issue~\cite{pcs22, crc}, resulting in worse coding performance than residual coding. This is most obvious when the prediction path that traverses from $x_c$ to the decoder output is lossy and the temporal prediction is nearly perfect, i.e. $x_c \approx x_t$. To alleviate the bottleneck issue, Brand et al.~\cite{crc} propose conditional residual coding, which encodes the prediction residue $x_t-x_c$ with a conditional inter-frame codec. In both lossless and lossy coding scenarios, conditional residual coding is shown to be at least as effective as conditional coding. However, this conclusion is valid under the assumption that the temporal predictor has good quality and the entropy of the residue $x_t-x_c$ is lower than that of the input frame $x_t$. Recognizing that these assumptions can be violated in regions with unreliable motion estimates or dis-occlusion, Chen et al.~\cite{maskcrt} propose masked conditional residual coding, where a soft mask is used to switch between conditional coding and conditional residual coding at pixel level. This work validates the hybrid buffering strategy in a masked conditional residual coding framework for its better coding efficiency. However, we stress that it is equally applicable to the other coding frameworks.
\section{Hybrid Temporal Information Propagation}
\label{sec:method}
\subsection{System Overview}
\label{sec:overview}
\begin{figure*}[t!]
    \centering
    \includegraphics[width=0.95\linewidth, trim=0 12 0 10, clip]{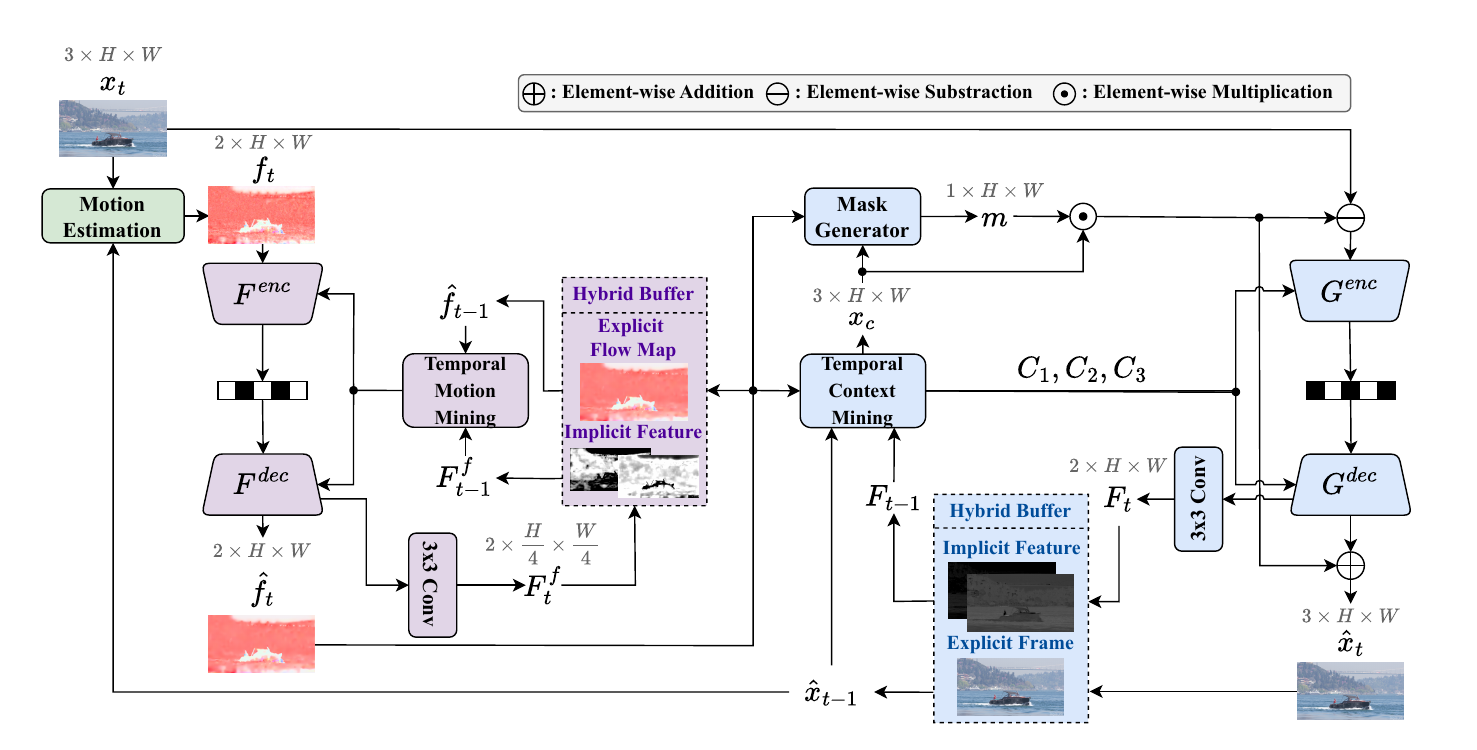}
    \vspace{-0.2cm}
    \caption{System overview of our proposed HyTIP. 
    The architectures of the temporal context mining module in the inter-frame coding, the inter-frame codec $\{G^{enc}, G^{dec}\}$, and the motion codec $\{F^{enc}, F^{dec}\}$ are adapted from \cite{fm}, while the mask generator follows the architecture of \cite{maskcrt}. Further architectural details are provided in the supplementary material.}
    \vspace{-0.4cm}
    \label{fig:overview}
\end{figure*}
Fig.~\ref{fig:overview} illustrates our HyTIP framework. It is a frame-based temporal predictive coding framework. (1) First, the motion estimation module (in green) performs motion estimation between an input frame $x_t$ and its reference frame $\hat{x}_{t-1}$ to obtain an optical flow map $f_t$. (2) Second, the motion coding modules (in purple) encode $f_t$ as $\hat{f}_t$. (3) Third, the decoded optical flow map $\hat{f}_t$ and the propagated temporal information in the buffer are used to generate temporal predictors for $x_t$ to assist in inter-frame coding (the blue modules). The process involves the pixel-domain predictor $x_c$ and the multi-scale feature-domain predictors $\{C_1, C_2, C_3\}$. (4) Finally, the temporal information is buffered and propagated for coding the next input frame. 

This frame-by-frame processing, along with the propagation of useful information to the next frame, can be viewed as an RNN operating along the temporal dimension. Based on how the buffered temporal information is used to generate temporal predictors, the mainstream inter-frame coding can be broadly classified into two major approaches, which implement the output-recurrence and hidden-to-hidden connections, respectively. The output-recurrence methods typically store the previous decoded frame $\hat{x}_{t-1}$ in an explicit buffer as the reference temporal information. On the other hand, the hidden-to-hidden methods buffer the latent features (i.e.~$F_{t-1}$) from the coding process for the previous frame $x_{t-1}$ in an implicit buffer. In this work, we integrate both the output-recurrence and hidden-to-hidden mechanisms within a unified coding framework and study the trade-offs between the coding performance and buffer size. In Fig.~\ref{fig:overview}, we use a $3 \times 3$ convolutional layer to adjust the channel size of the latent features stored in the implicit buffer. Additionally we explore how various buffering strategies for motion coding impact the rate-distortion-complexity trade-offs.

Our inter-frame codec adheres to the masked conditional residual coding framework~\cite{maskcrt}. The input signal to the conditional inter-frame codec is $x_t - m \odot x_c$, where the multi-scale feature-domain temporal predictors $\{C_1, C_2, C_3\}$ serves as condition signals. Here, $\odot$ denotes element-wise multiplication, and $m$ is a pixel-wise soft mask generated using the decoded flow map $\hat{f}_{t-1}$ and the pixel-domain temporal predictor $x_c$. The values of $m$ range from 0 to 1, with all three channels of $m$ sharing the same mask value. Since the input signal $ x_t - m \odot x_c $ is equivalent to $(1-m) \odot x_t + m \odot (x_t - x_c)$, our inter-frame codec forms a hybrid of conditional coding and conditional residual coding. Notably, we \textcolor{black}{adapt} the CNN-based network structure from~\cite{fm}, instead of \textcolor{black}{using} the Transformer-based coding backbone from \cite{maskcrt}. More details are provided in the supplementary document.

\subsection{Inter-frame Coding with Hybrid Buffering}
\label{sec:inter}
Our HyTIP employs a hybrid buffering strategy that dedicates a hybrid buffer to storing both the previously decoded frame $\hat{x}_{t-1}$ and the latent features  $F_{t-1}$ for temporal prediction. As depicted in Fig.~\ref{fig:overview}, these references are motion compensated with the decoded flow map $\hat{f}_t$, generating the pixel-domain temporal predictor $x_c$ and the multi-scale features $\{C_1,C_2, C_3\}$. The predictor $x_c$ forms a masked temporal prediction of the input frame $x_t$, with the prediction residues passed through the contextual encoder $G^{enc}$ and decoder $G^{dec}$. Both $G^{enc}$ and $G^{dec}$ are conditioned on the same signals $\{C_1,C_2, C_3\}$.

In our scheme, the previously decoded frame $\hat{x}_{t-1}$ serves as the primary, but not the sole, source of temporal information for generating the temporal predictor and condition signals. The additional latent features $F_{t-1}$ provide complementary information to enhance coding efficiency. When viewed as a form of RNN, our inter-frame codec features both the output-recurrence and hidden-to-hidden connections. It avoids the dual constraint issue, which requires $\hat{x}_{t-1}$ to be a good approximation of $x_{t-1}$ while carrying sufficient information from the past to encode $x_t$. Compared to the schemes that rely solely on the hidden-to-hidden connection, where $F_{t-1}$ is the sole source of temporal information and is learned completely from data, our hybrid buffer leverages the prior knowledge that the previously decoded frame $\hat{x}_{t-1}$ correlates highly with the current input frame $x_t$ to reduce the reliance on the latent features $F_{t-1}$, thereby reducing the size of the implicit buffer. The net effect is a significant reduction in buffer size while maintaining comparable or even better coding efficiency. It is worth noting that most hidden-to-hidden-only methods inevitably buffer $\hat{x}_{t-1}$ for motion estimation, in order to propagate the latent features along the temporal dimension. However, the generation of their coded latents for $x_t$ does not direct use $\hat{x}_{t-1}$. This design feature is in direct contrast to our scheme.

\subsection{Motion Coding with Hybrid Buffering}
\label{sec:motion}
Following the same hybrid buffering strategy as our inter-frame coding, we extend this idea to motion coding, the task of which is to encode optical flow maps. Unlike video frames, which typically contain many high-frequency details, optical flow maps are generally smoother signals along the spatial dimension, with mostly low-frequency components. Oftentimes, the optical flow evolves slowly over time in typical video sequences. Therefore, we buffer and propagate the decoded flow map $\hat{f}_{t-1}$ as an explicit temporal reference, and also the latent features $F^f_{t-1}$ of the motion coding process for $f_{t-1}$ as an implicit temporal reference. It is worth noting that these temporal references are used for coding $f_t$ without motion compensation. In a manner similar to~\cite{dc, fm}, the latent features $F^f_{t-1}$ are stored at one-fourth of the input resolution in both width and height. However, unlike \cite{dc, fm}, which adopts a hidden-to-hidden connection approach by buffering only $F^f_{t-1}$, our method adopts a hybrid buffering strategy that stores both $\hat{f}_{t-1}$ and $F^f_{t-1}$ as temporal references.

\section{Experiments}
\label{sec:experiment}
\subsection{Experimental Setup}
\noindent\textbf{Training details:} \textcolor{black}{We train our models with 5-frame training first on the Vimeo-90K dataset~\cite{vimeo}, consisting of 91,701 7-frame sequences, and fine-tune our models with 10-frame training on BVI-DVC~\cite{bvi}, consisting of 800 64-frame sequences.} \new{To optimize our variable-rate model for PSNR-RGB, the hyperparameter $\lambda$, which balances rate and distortion in the objective function, is randomly sampled from [227, 2032]. Similarly, for the model optimized for MS-SSIM, $\lambda$ is sampled from [7, 46]. Additional training details are provided in the supplementary material.} 

\noindent\textbf{Evaluation Methodologies:} We evaluate our method on the UVG~\cite{uvg}, MCL-JCV~\cite{mcl}, and HEVC Class B-E~\cite{hevcctc}, and HEVC-RGB~\cite{hevcrgb} datasets. \textcolor{black}{Following the common test protocol for learned video coding, we convert the test sequences from YUV420 to RGB444 using BT.601. For BT.709, which is only used by DCVC-DC and DCVC-FM, we report the results in the supplementary materials.} For each test sequence, the first 96 frames are encoded, and the intra period is set to 32. To ensure a fair comparison and avoid padding, the width and height of all video frames are cropped to multiples of 64. Intra coding is applied at scene cuts for all the competing methods. We report PSNR \textcolor{black}{and MS-SSIM} in the RGB domain and birate in bit-per-pixel (bpp). Following the common test protocol for learned video codecs, the dataset BD-rate is reported; in computing the BD-rate, a dataset-specific rate-distortion point is computed by averaging the per-frame PSNR-RGB \textcolor{black}{(or MS-SSIM-RGB)} and bits-per-pixel across all coded frames in the dataset.
Positive and negative BD-rate numbers indicate rate inflation and reduction, respectively.
\begin{table*}[t]
\fontsize{9pt}{9pt}\selectfont
\caption{BD-rate (\%) comparison of different buffering strategies with different buffer sizes. The anchor employs explicit buffering in both motion and inter-frame coding. \new{The values in parentheses indicate the number of full-resolution feature maps buffered for coding one input flow map or frame (explicit + implicit). One flow map and one RGB frame are equivalent to two and three full-resolution feature maps, respectively. For clarity, buffer sizes for entropy coding are excluded here, in contrast to Section~\ref{sec:sota}.} 
}
\vspace{-0.2cm}
\centering
\label{table:buffer_size}
\setlength{\tabcolsep}{5pt}
\begin{tabular}{l|l|rrrrrrr|r}
\toprule
Motion            & Inter         & UVG   & MCL-JCV & HEVC-B & HEVC-C & HEVC-D & HEVC-E & HEVC-RGB & Average \\
\midrule
Explicit (2+0)      & Explicit (3+0)  & 0     & 0       & 0      & 0      & 0      & 0      & 0        & 0       \\
\midrule
Implicit (0+4)      & Explicit (3+0)  & -11.5 & -9.5    & -12.6  & -18.6  & -12.6  & -11.9  & -8.8     & -12.2   \\
Implicit (0+2.125)  & Explicit (3+0)  &  -6.5 & -3.8    &  -6.2  & -10.4  &  -6.7  &  -8.6  & -3.0     &  -6.5   \\
\midrule
Hybrid (2+4)        & Explicit (3+0)  & -15.1 & -13.2   & -15.4  & -23.8  & -19.9  & -15.8  &  -9.1    & -16.0   \\
Hybrid (2+0.125)    & Explicit (3+0)  & -13.0 & -11.2   & -14.1  & -19.1  & -16.3  & -18.1  & -11.0    & -14.7   \\
\midrule
Hybrid (2+0.125)    & Implicit (0+51) & -12.6 & -15.5   & -20.1  & -30.2  & -27.8  & -20.6  & -10.5    & -19.6   \\
Hybrid (2+0.125)    & Implicit (0+5)  & -9.2  & -13.0   & -14.9  & -24.6  & -21.4  & -14.6  &  -7.4    & -15.0   \\
\midrule
Hybrid (2+0.125)    & Hybrid (3+48)   & -17.3 & -16.3   & -21.0  & -30.3  & -27.3  & -25.7  & -15.1    & -21.9   \\
Hybrid (2+0.125)    & Hybrid (3+2)    & -17.6 & -15.9   & -20.3  & -29.1  & -25.9  & -26.6  & -14.8    & -21.5   \\
\bottomrule
\end{tabular}
\vspace{-0.2cm}
\end{table*}
\begin{table*}[t]
\fontsize{9pt}{9pt}\selectfont
\caption {BD-rate (\%) comparison of longer sequence training impact on three buffering strategies for inter-frame coding. The anchor is the variant employing explicit buffering in both motion and inter-frame coding. \new{The values in parentheses, as in Table~\ref{table:buffer_size}, indicate the number of full-resolution feature maps buffered for coding one input flow map or frame (explicit + implicit).} 
}
\vspace{-0.2cm}
\centering
\label{table:long_sequence}
\setlength{\tabcolsep}{3.5pt}
\begin{tabular}{l|l|r|rrrrrrr|r}
\toprule
Motion                            & Inter                         & \# Frame & UVG   & MCL-JCV & HEVC-B & HEVC-C & HEVC-D & HEVC-E & HEVC-RGB & Average \\ 
\midrule
\multirow{2}{*}{Hybrid   (2+0.125)} & \multirow{2}{*}{Explicit (3+0)} & 5        & -13.0 & -11.2   & -14.1  & -19.1  & -16.3  & -18.1  & -11.0    & -14.7   \\ 
                                  &                               & 10       & -19.9 & -16.3   & -16.1  & -20.9  & -18.4  & -20.2  & -13.2    & -17.9   \\ 
\midrule
\multirow{2}{*}{Hybrid (2+0.125)}   & \multirow{2}{*}{Implicit (0+5)} & 5        & -9.2  & -13.0   & -14.9  & -24.6  & -21.4  & -14.6  & -7.4     & -15.0   \\ 
                                  &                               & 10       & -21.4 & -22.0   & -21.0  & -30.2  & -25.2  & -13.2  & -12.6    & -20.8   \\ 
\midrule
\multirow{2}{*}{Hybrid (2+0.125)}   & \multirow{2}{*}{Hybrid (2+3)}   & 5        & -17.6 & -15.9   & -20.3  & -29.1  & -25.9  & -26.7  & -14.8    & -21.5   \\ 
                                  &                               & 10       & -25.4 & -21.7   & -25.1  & -34.5  & -30.6  & -29.1  & -20.3    & -26.7   \\ 

\bottomrule
\end{tabular}
\vspace{-0.3cm}
\end{table*}

\subsection{Experimental Results}
\label{sec:main_exp}
This section compares our hybrid buffering strategy with the single use of the explicit or implicit strategy. For a fair assessment, we implement these buffering strategies with the same coding framework (Fig.~\ref{fig:overview}). Moreover, we align the buffer size for the hybrid and implicit strategies. As an example, if the hybrid variant keeps one previously decoded frame (3 channels) and two latent feature maps, then the implicit variant maintains five latent feature maps. \textcolor{black}{To reduce the training time required for extensive experiments, we disable the hierarchical quality structure~\cite{fm}, the channel transform module~\cite{maskcrt}, the checkerboard context model~\cite{multistage} in the inter-frame codec, \new{and the variable-rate modules}, as opposed to the full model used in Section~\ref{sec:sota}. More architectural details are provided in the supplementary material. }


\noindent \textbf{Explicit, Implicit, and Hybrid Buffering:} 
Table~\ref{table:buffer_size} compares the coding performance of adopting various buffering strategies for motion and inter-frame coding. For this experiment, we train the competing methods on 5-frame training sequences in Vimeo-90K only. From Table~\ref{table:buffer_size}, we make the following observations: 

(1) \textit{Robustness of propagating latent features with a large buffer}: As expected, both the hybrid and implicit variants with a large buffer size outperform the explicit one in both motion and inter-frame coding, as they can propagate a larger number of latent feature maps, using more effectively temporal information. 

(2) \textit{Sensitivity of implicit buffering to buffer size reduction:} The coding performance of the implicit buffering strategy, when applied to motion or inter-frame coding, decreases significantly with reduced buffer size. With the implicit strategy, the model solely relies on learning, in order to extract useful temporal information. It is challenging to train a model that is capable of extracting both useful and compact temporal features.

\begin{figure}[t!]
    \centering
        \begin{subfigure}{1\linewidth}
        \centering
        \includegraphics[width=\linewidth, trim= 10 10 10 10, clip]{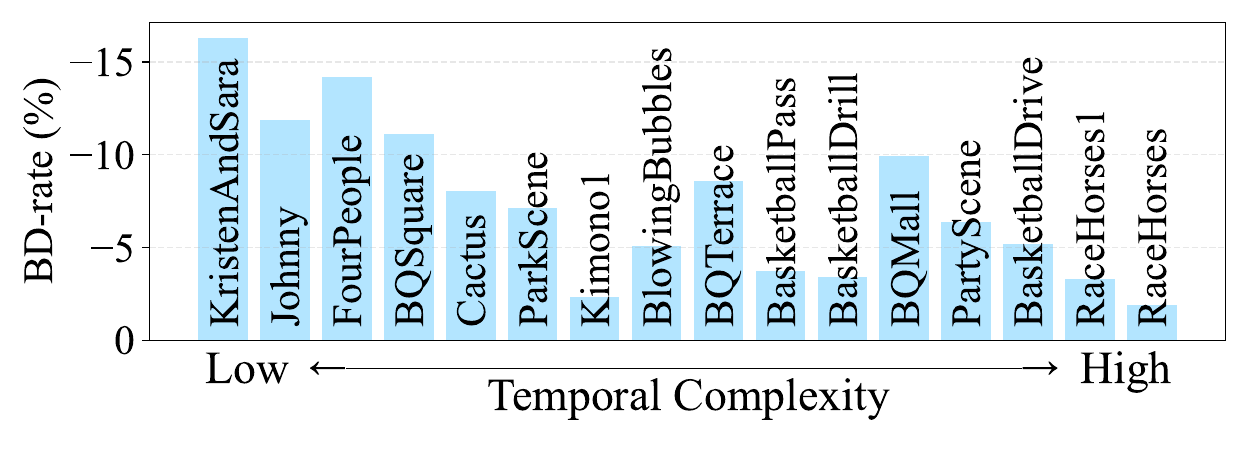}
    \end{subfigure}
    \vspace{-0.7cm}
    \caption[Per-sequence comparison between hybrid and implicit buffering strategies.]{\new{Per-sequence BD-rate results of the inter-frame codec using hybrid versus implicit buffering (anchor) on the HEVC Class B-E datasets, both under the small buffer setting in Table~\ref{table:buffer_size}. Results are ordered by sequence temporal complexity~\cite{vca}.}}
    \vspace{-0.5cm}
    \label{fig:per_seq}
\end{figure}
\begin{table}[t]
\fontsize{9pt}{9pt}\selectfont
\caption{\new{Complexity comparison of encoding/decoding MACs and model size across buffering strategies.} \new{The values in parentheses, as in Table~\ref{table:buffer_size}, indicate the number of full-resolution feature maps buffered for coding one input flow map or frame (explicit + implicit).}}
\vspace{-0.2cm}
\label{tab:complexity_ablation12}
\centering
\setlength{\tabcolsep}{4.2pt}
\begin{tabular}{l|l|ccc}
\toprule
     \begin{tabular}[c]{@{}c@{}} Motion  \end{tabular} 
    & \begin{tabular}[c]{@{}c@{}} Inter  \end{tabular} 
    & \begin{tabular}[c]{@{}c@{}} Model \\ Size (M) \end{tabular} 
    & \begin{tabular}[c]{@{}c@{}} Enc. / Dec. \\ kMACs/pixel \end{tabular}  \\
\midrule
Explicit (2+0)     & Explicit (3+0)  & 13.712 & 1277 / 858  \\
\midrule
Implicit (0+4)     & Explicit (3+0)  & 13.358 & 1268 / 849 \\
Implicit (2+0.125) & Explicit (3+0)  & 13.323 & 1266 / 847 \\
\midrule
Hybrid (2+4)       & Explicit (3+0)  & 13.877 & 1284 / 865 \\
Hybrid (2+0.125)   & Explicit (3+0)  & 13.805 & 1280 / 861 \\
\midrule
Hybrid (2+0.125)   & Implicit (0+51) & 14.918 & 1303 / 884 \\
Hybrid (2+0.125)   & Implicit (0+5)  & 14.891 & 1281 / 862 \\
\midrule
Hybrid (2+0.125)   & Hybrid (3+48)   & 14.917 & 1302 / 883 \\
Hybrid (2+0.125)   & Hybrid (2+3)    & 14.890 & 1280 / 861 \\
\bottomrule
\end{tabular}
\vspace{-0.4cm}
\end{table}

(3) \textit{Resilience of hybrid buffering to buffer size reduction:} In contrast, our hybrid buffering strategy with small buffer size performs comparable to the large buffer size variant, 
with a reduced buffer size for latent features in either motion or inter-frame codec. Unlike the implicit strategy, our hybrid strategy relies heavily on the previously decoded frame as the primary source for temporal prediction. The latent features play a secondary role in complementing the previously decoded frame, which generally has the highest correlation with the current input frame.

\begin{table*}[t]
\fontsize{9pt}{9pt}\selectfont
\caption{BD-rate (\%) comparison between HyTIP and the state-of-the-art methods in terms of PSNR-RGB. The anchor is VTM 17.0. 
}
\vspace{-0.2cm}
\label{tab:sota_psnr_gop32}
\centering
\begin{tabular}{l|rrrrrrr|r}
\toprule
                        & UVG    & MCL-JCV & HEVC-B & HEVC-C & HEVC-D & HEVC-E & HEVC-RGB & Average \\
\midrule
VTM 17.0~\cite{vtm}     &   0.0  & 0.0     & 0.0    & 0.0    & 0.0    & 0.0    & 0.0      & 0.0     \\
HM 16.25~\cite{hm}      &  26.3  & 36.8    & 31.8   & 29.8   & 29.6   & 32.5   & 34.0     & 31.5    \\
MaskCRT~\cite{maskcrt}  & -10.1  & 7.4     & 2.7    & 21.6   & -6.4   & 17.1   & -3.4     & 4.1     \\
DCVC-TCM~\cite{tcm}     &  16.1  & 27.4    & 29.3   & 59.8   & 18.8   & 61.3   & 24.3     & 33.9    \\
DCVC-HEM~\cite{hem}     & -19.2  & -8.5    & -4.4   & 14.9   & -15.0  & 2.5    & -11.0    & -5.8    \\
DCVC-DC~\cite{dc}       & -29.9  & -21.4   & -16.5  & -9.4   & -30.3  & -28.1  & -29.7    & -23.6   \\
DCVC-FM~\cite{fm}       & -23.9  & -13.4   & -10.9  & -5.4   & -26.9  & -29.2  & -19.7    & -18.5   \\
HyTIP (Ours)       & -34.7  & -25.0   & -23.7  & -16.4  & -35.6  & -21.2  & -29.0    & -26.5   \\  
\bottomrule
\end{tabular}
\label{tab:SOTA_RD-PSNR_RGB_I32_BT601}
\vspace{-0.2cm}
\end{table*}

\begin{table*}[t]
\fontsize{9pt}{9pt}\selectfont
\caption{BD-rate (\%) comparison between HyTIP and the state-of-the-art methods in terms of MS-SSIM-RGB. The anchor is VTM 17.0. 
}
\vspace{-0.2cm}
\label{tab:sota_ssim_gop32}
\centering
\begin{tabular}{l|rrrrrrr|r}
\toprule
                         & UVG   & MCL-JCV & HEVC-B & HEVC-C & HEVC-D & HEVC-E & HEVC-RGB & Average \\
\midrule
VTM      ~\cite{vtm}     & 0.0   & 0.0     & 0.0    & 0.0    & 0.0    & 0.0    & 0.0      & 0.0     \\
HM       ~\cite{hm}      & 20.2  & 30.2    & 27.4   & 27.8   & 29.0   & 28.8   & 27.2     & 27.2    \\
MaskCRT  ~\cite{maskcrt} & -27.0 & -36.5   & -44.9  & -34.1  & -48.5  & -39.3  & -42.9    & -39.0   \\
DCVC-TCM ~\cite{tcm}     & -12.1 & -25.3   & -31.3  & -21.6  & -38.8  & -29.9  & -25.9    & -26.4   \\
DCVC-HEM ~\cite{hem}     & -31.5 & -45.0   & -50.6  & -43.3  & -57.5  & -55.7  & -45.6    & -47.0   \\
DCVC-DC  ~\cite{dc}      & -37.3 & -50.9   & -56.8  & -54.4  & -64.9  & -66.0  & -55.8    & -55.2   \\
HyTIP (Ours)             & -42.8 & -53.4   & -59.7  & -54.8  & -64.4  & -64.0  & -57.8    & -56.7   \\  
\bottomrule
\end{tabular}
\label{tab:SOTA_RD-MSSSIM_RGB_I32_BT601}
\vspace{-0.5cm}
\end{table*}

(4) \textit{Superiority of hybrid buffering with a small buffer:} Compared to the explicit strategy, our hybrid strategy with a small buffer size achieves notable gains in both motion and inter-frame coding (\textcolor{black}{14.7\%} and additional \textcolor{black}{6.8\%} BD-rate savings, respectively). It is to be noted that the hybrid buffering uses a slightly larger buffer size for motion and inter-frame coding (2 additional feature maps of one-fourth the input spatial resolution, \textcolor{black}{which is equivalent to 0.125 channel of the input resolution feature,} for motion coding, and likewise, 2 additional feature maps of the input resolution for inter-frame coding). The explicit strategy underperforms due mainly to the dual constraint issue (see Fig.~\ref{fig:teaser}), where the decoded frame must meet two requirements: (1) closely approximating the input frame and (2) serving as a sufficient summary of past information for propagation. In comparison, the implicit strategy with a small buffer size shows some improvement over the explicit one (\textcolor{black}{6.5\%} and additional \textcolor{black}{0.3\%} BD-rate savings, respectively). However, its gain is not as significant as that of our hybrid approach due to its purely learning-based nature, which fails to take advantage of the prior knowledge that the previously decoded frame correlates highly with the current input frame.

\new{Fig.~\ref{fig:per_seq} further presents the per-sequence BD-rate savings of hybrid buffering strategy to the implicit buffering strategy. As shown, the hybrid strategy has greater gains than the implicit strategy in sequences with lower temporal complexity. This is likely because the hybrid strategy allows the model to directly use the previous decoded frame, which typically has the highest correlation with the current frame.} 



\vspace{0.1cm}
\noindent\textbf{Training with longer sequences:} Generally, training with longer sequences can enhance the learning of RNNs by allowing them to capture long-term dependencies. Table~\ref{table:long_sequence} analyzes the impact of long-sequence training on the three buffering strategies in the context of inter-frame coding. Specifically, we fine-tune each initial codec, originally trained on Vimeo-90K~\cite{vimeo} with 5-frame sequences, using BVI-DVC~\cite{bvi} with 10-frame sequences. From Table~\ref{table:long_sequence}, all three approaches benefit from long-sequence training. However, the explicit buffering strategy achieves a performance gain of only 3.2\%, which is lower than the 5.8\% and 5.2\% gains observed in the implicit and hybrid buffering strategies, respectively. This discrepancy arises from the dual constraint on the output frame in explicit buffering. In contrast, the implicit and hybrid buffering strategies, which propagate implicit features without such constraints, are better able to leverage long-sequence training.

\vspace{0.1cm}

\noindent\textbf{Complexity comparison:} Table~\ref{tab:complexity_ablation12} presents the complexity comparison of various buffering strategies in terms of the model size, kilo-multiply-accumulate operations per pixel (kMACs/pixel), and buffer size. \new{As reported in Table~\ref{table:buffer_size}, our hybrid scheme has the best coding performance under a similar buffer size, while Table~\ref{tab:complexity_ablation12} further confirms that its model size and MAC are comparable to the other schemes.} The implicit and hybrid buffering strategies introduce only two minor structural modifications compared to the explicit buffering strategy: (1) an additional CNN to adjust the channel size of the buffered implicit features in the motion and inter-frame codecs and (2) a slight change to the first-layer convolution in temporal context mining (or temporal motion mining) for generating inter-frame (or motion) temporal predictors (see Fig.~\ref{fig:overview}). As a result, the increase in the model size and encoding/decoding kMAC/pixel is negligible. Although our hybrid buffering strategy with a small buffer size requires storing additionally \textcolor{black}{0.125} and \textcolor{black}{2} channels of full-resolution implicit features for motion and inter-frame coding, respectively, compared to the explicit buffering strategy, it still requires significantly fewer channels than the 48+ implicit features used in state-of-the-art methods~\cite{tcm,hem,dc,fm}. To put things into context, traditional codecs typically buffer four reference frames.


\subsection{Comparison with the State-of-the-Arts}
\label{sec:sota}
This section compares our method with state-of-the-art traditional codecs and learned P-frame codecs, in terms of rate-distortion performance. The traditional codecs include HM 16.25~\cite{hm} and VTM 17.0~\cite{vtm} under the low-delay-B configuration. Following \cite{dc}, both HM and VTM are configured to encode the input video in YUV444 as the internal color space. This setting is shown to achieve better coding results than using YUV420 when the final distortion is measured in the RGB domain. The learned P-frame codecs include DCVC-TCM~\cite{tcm}, DCVC-HEM~\cite{hem}, DCVC-DC~\cite{dc}, and DCVC-FM~\cite{fm}, which adopt the implicit buffering strategy, as well as MaskCRT~\cite{maskcrt}, which employs the explicit buffering strategy. 

Tables~\ref{tab:sota_psnr_gop32} and~\ref{tab:sota_ssim_gop32} present the coding performance of the competing methods in terms of PSNR and MS-SSIM, respectively. As shown, our HyTIP achieves comparable coding performance to the state-of-the-art method DCVC-DC~\cite{dc}, except on HEVC-E, which is a special dataset that consists of video conferencing-type sequences with static backgrounds. Aside from DCVC-DC, HyTIP outperforms the other methods, including traditional codecs HM 16.25~\cite{hm} and VTM 17.0~\cite{vtm}. 

It is important to note that the training procedure significantly impacts the coding performance of a neural video codec. Since the state-of-the-art DCVC-DC~\cite{dc} and DCVC-FM~\cite{fm} have not released their training details, we are unable to reproduce their results for a fair comparison. The results shown here are obtained with their publicly available test models. However, this work adopts the training procedure from \cite{maskcrt}, which is different from those for DCVC-DC~\cite{dc} and DCVC-FM~\cite{fm}. Consequently, even though our HyTIP achieves performance similar to DCVC-DC~\cite{dc} or slightly worse on HEVC-E, this does not imply that our hybrid buffering strategy is ineffective. According to the experiments in Section~\ref{sec:main_exp}, there is still potential to combine their approaches with ours to achieve further gains.

\begin{table}[t]
\fontsize{9pt}{9pt}\selectfont
\caption{Complexity comparison between HyTIP and the state-of-the-art learned P-frame codecs. 
Buffer size represents the number of full-resolution feature maps that need to be buffered for coding one input frame.}
\vspace{-0.2cm}
\label{tab:complexity_sota}
\centering
\setlength{\tabcolsep}{5pt}
\begin{tabular}{l|ccc}
\toprule
      \begin{tabular}[c]{@{}c@{}} Methods                  \end{tabular} 
    & \begin{tabular}[c]{@{}c@{}} Model \\ Size (M)        \end{tabular} 
    & \begin{tabular}[c]{@{}c@{}} Enc. / Dec. \\ kMACs/pixel \end{tabular}
    & \begin{tabular}[c]{@{}c@{}} Buffer \\ Size           \end{tabular}  \\
\midrule
DCVC-TCM~\cite{tcm}    &  10.709  &  1406 / \; 917  &  67      \\
DCVC-HEM~\cite{hem}    &  17.523  &  1662 /   1243  &  67.625  \\
DCVC-DC~\cite{dc}      &  19.779  &  1343 / \; 918  &  \new{55.75}   \\
DCVC-FM~\cite{fm}      &  18.336  &  1137 / \; 871  &  \new{55.75}   \\
MaskCRT~\cite{maskcrt} &  31.152  &  1401 / \; 763  &  13      \\
HyTIP (Ours)           &  16.593  &  1293 / \; 873  &   7.875  \\   
\bottomrule
\end{tabular}
\vspace{-0.4cm}
\end{table}

\begin{figure}[t]
    \centering
    \includegraphics[width=0.99\linewidth]{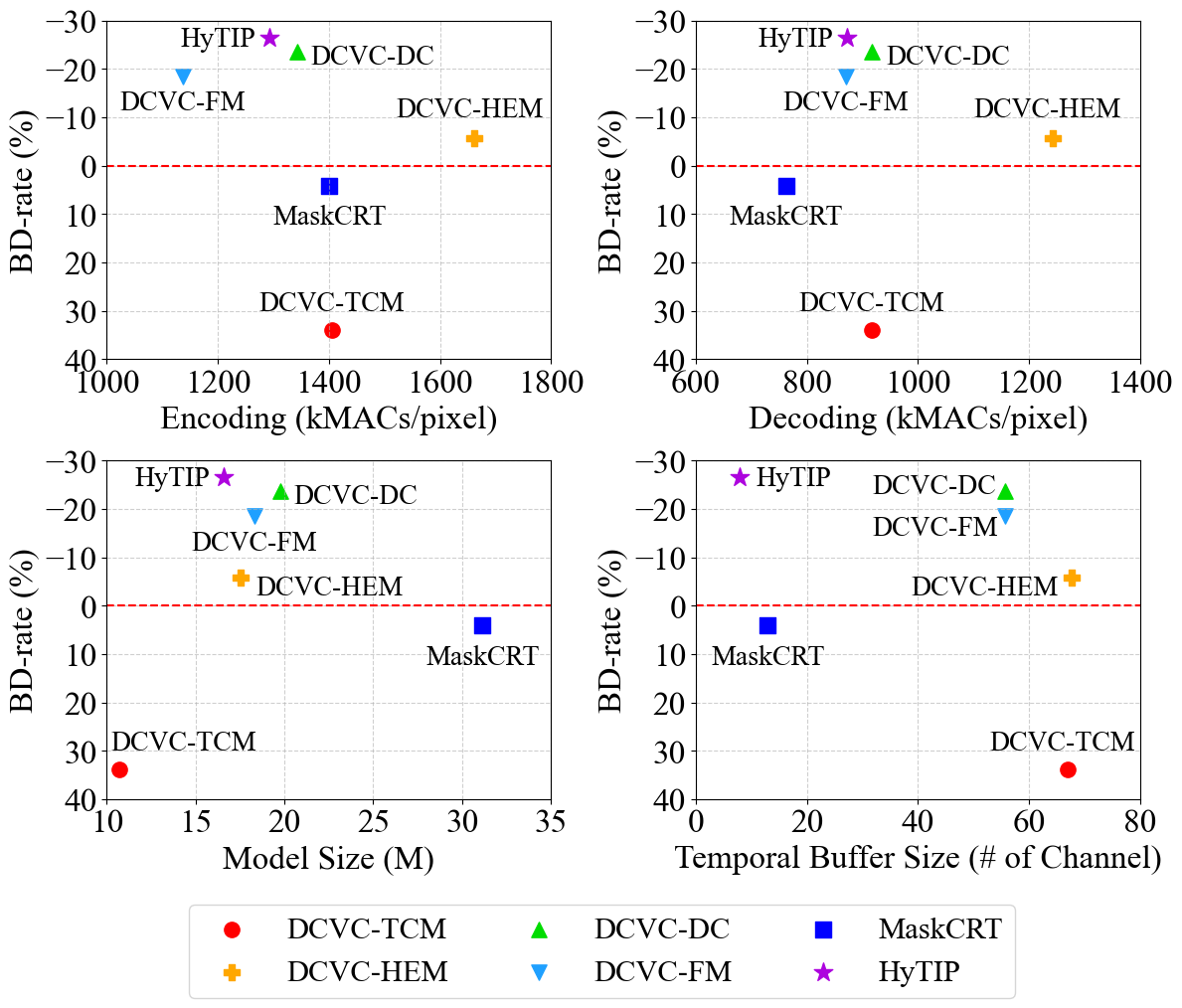}
    \vspace{-0.2cm}
    \caption[Characterization of complexity-performance trade-offs of our HyTIP and the state-of-the-art methods.]{\new{Comparison of complexity-performance trade-offs between HyTIP and the state-of-the-art methods. 
    The vertical axis is the BD-rate savings in terms of PSNR-RGB evaluated with VTM-17.0 (Low-delay B) serving as the anchor. Positive and negative BD-rate numbers indicate rate inflation and reduction, respectively. The horizontal axes are the complexity metrics.}}
    \label{fig:complexity_visual_sota}
    \vspace{-0.5cm}
\end{figure}

Table~\ref{tab:complexity_sota} compares the complexity of the competing methods in terms of the model size, kMACs/pixel, and buffer size. \new{Fig.~\ref{fig:complexity_visual_sota} reports how these methods trade off between complexity and BD-rate.} As shown, HyTIP requires the smallest buffer size among all the competing methods, utilizing only 14\% of the buffer size required by works using the implicit buffering strategy, such as DCVC-TCM~\cite{tcm}, DCVC-HEM~\cite{hem}, DCVC-DC~\cite{dc}, and DCVC-FM~\cite{fm}. Additionally, the decoding kMAC/pixel of HyTIP is comparable to DCVC-FM~\cite{fm} and slightly lower than DCVC-DC~\cite{dc}. \new{Note that buffer size reflects the amount of data that must be fetched at each time step for encoding/decoding a video frame. Typically, the buffer stores previously decoded frames and/or features for temporal prediction resides in off-chip memory. A larger buffer size in this context requires fetching more data from off-chip memory, leading to higher memory bandwidth demands and reduced practicality. A similar design constraint was adopted in traditional video codecs, where block prediction typically relies on only two reference blocks, motivates our focus on reducing buffer size.} While large buffer size is intrinsic to the algorithm design, both the model size and kMAC can be further reduced through network optimization.

\section{Conclusion}
\label{sec:conclusion}
In this work, we revisit how typical learned video compression frameworks propagate temporal information to assist coding from an RNN perspective, and propose HyTIP, combining output-recurrence and hidden-to-hidden connections to leverage the advantages of both approaches. By propagating the previously decoded frame as explicit information to primarily serve as temporal data, we only require a small number of implicit latent features to carry complementary temporal information, achieving competitive performance. Our experimental results confirm the superiority of the combined use of implicit and explicit buffering strategies over the use of either alone. This hybrid approach is less sensitive to buffer size than purely implicit buffering strategies. Compared to state-of-the-art methods that adopt solely implicit buffering strategy, our HyTIP requires only 14\% of the buffer size while achieving comparable performance on most testsets. In this work, we aim to investigate the design of temporal propagation mechanisms within learned video codecs, rather than their functionality. Extending our codec to enhance its functionality, such as \new{YUV coding}, is among our future directions.

\vspace{-0.4cm}
\new{\section*{Acknowledgement}}
\noindent \new{This work is supported by MediaTek and National Science and Technology Council (NSTC), Taiwan, under Grants 113-2634-F-A49-007- and 111-2923-E-A49-007-MY3. We thank National Center for High-performance Computing (NCHC) for providing computational and storage resources, and NVAITC for providing access to the Taipei-1 supercomputer.}

{
    \small
    \bibliographystyle{ieeenat_fullname}
    \bibliography{main}
}

\clearpage
\beginsupplement







\maketitlesupplementary

\thispagestyle{empty}


%
This supplementary document provides the following additional materials and results to assist with the understanding of our HyTIP.

\begin{itemize}
\item \new{Additional results on buffer size} in Section~\ref{sec:buffer};
\item Results on long-sequence training under BT.709 color space conversion in Section~\ref{sec:long_709};
\item Rate-distortion comparisons to state-of-the-arts in Section~\ref{sec:SOTA_RD};
\item Network architecture details in Section~\ref{sec:architecture};
\item Training details in Section~\ref{sec:training_details};
\item Configurations of HM 16.25 and VTM 17.0 in Section~\ref{sec:configuration};

\end{itemize}
%
\section{Additional Results on Buffer Size}
\label{sec:buffer}
\new{Table~\ref{table:ablation_BT601} and Table~\ref{table:ablation_BT709} provide additional results on the buffer size configurations reported in Table~\ref{table:buffer_size} of the main paper, using BT.601 and BT.709 color space conversions, respectively. As shown, the conclusions are consistent with those in the main paper, where the hybrid buffering strategy demonstrates strong robustness and efficiency. It achieves better performance than both the explicit and implicit buffering strategies and exhibits greater resilience to buffer size reduction compared to the implicit buffering strategy. According to the results in Table~\ref{table:ablation_BT601} and Table~\ref{table:ablation_BT709}, our final design buffers 1 explicit decoded flow map and 0.125 full-resolution feature map (i.e., 2+0.125 full-resolution feature maps) for motion coding, and 1 explicit decoded frame and 2 full-resolution feature maps (i.e., 3+2 full-resolution feature maps) for inter-frame coding to balance coding performance and complexity.}


\section{Results on Long-sequence Training under BT.709 Color Space Conversion}
\label{sec:long_709}
\new{Table \ref{table:long_sequence_709} presents the results of the same study as Table~\ref{table:long_sequence} in the main paper, but employs BT.709 color space conversion, following \cite{dc, fm}, instead of BT.601 for the YUV420 to RGB444 conversion. As shown, the conclusions remain consistent with those in the main paper, where training with longer sequences generally improves all buffering strategies. The explicit buffering strategy shows a smaller performance gain due to its dual constraint on the output frame, while the implicit and hybrid strategies, which propagate implicit features without such constraints, are better able to leverage long-sequence training.}

\section{Rate-distortion Comparisons to State-of-the-arts}
\label{sec:SOTA_RD}

\begin{figure*}[h]
    \vspace{-2em}
    \begin{center}
    \begin{subfigure}{0.37\linewidth}
        \centering
        \includegraphics[width=\linewidth]{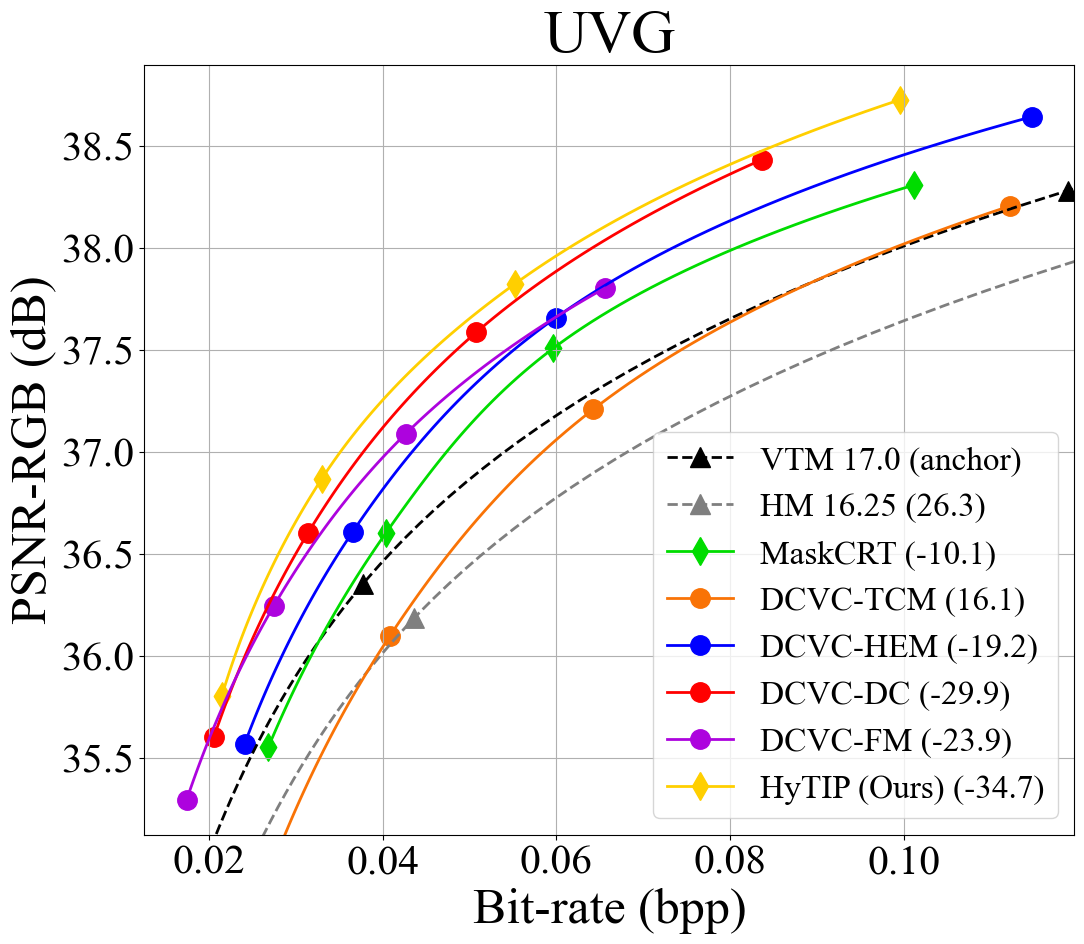}
    \end{subfigure}
    \begin{subfigure}{0.37\linewidth}
        \centering
        \includegraphics[width=\linewidth]{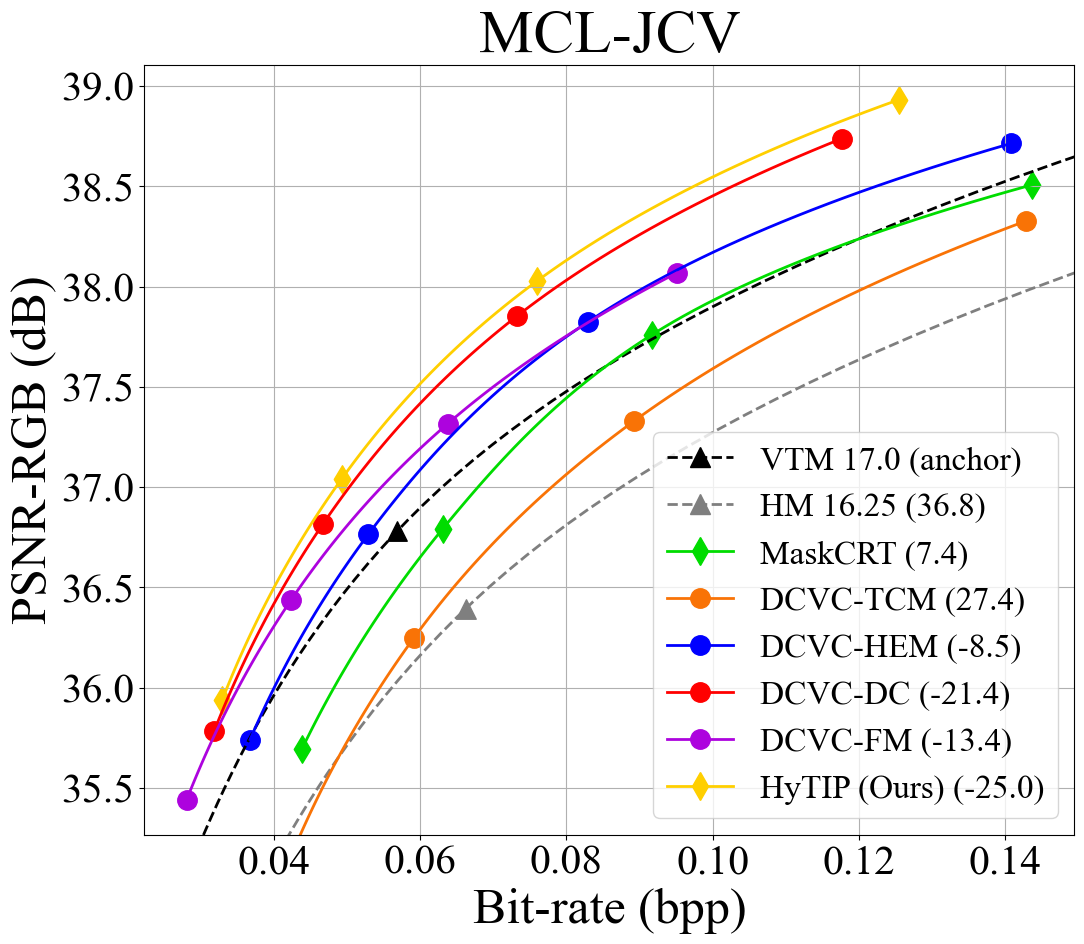}
    \end{subfigure}
    \begin{subfigure}{0.37\linewidth}
        \centering
        \includegraphics[width=\linewidth]{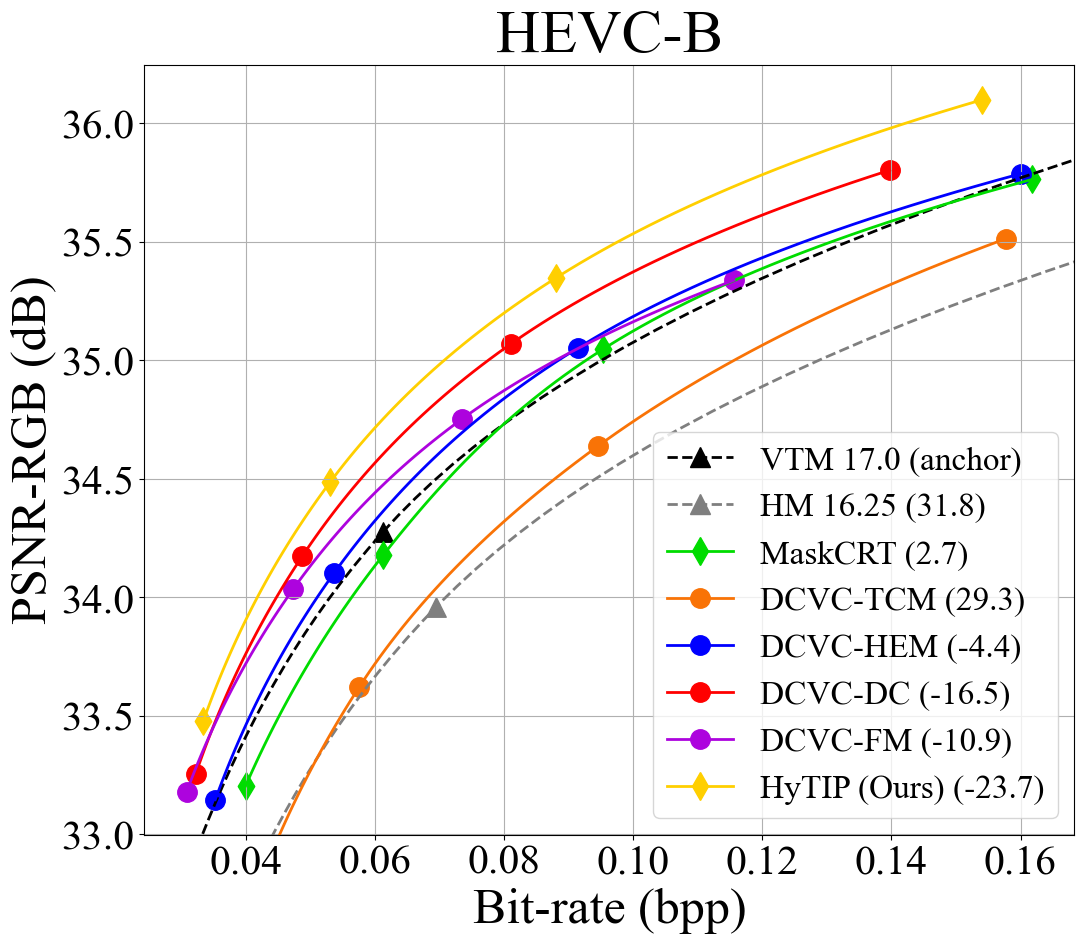}
    \end{subfigure}
    \begin{subfigure}{0.37\linewidth}
        \centering
        \includegraphics[width=\linewidth]{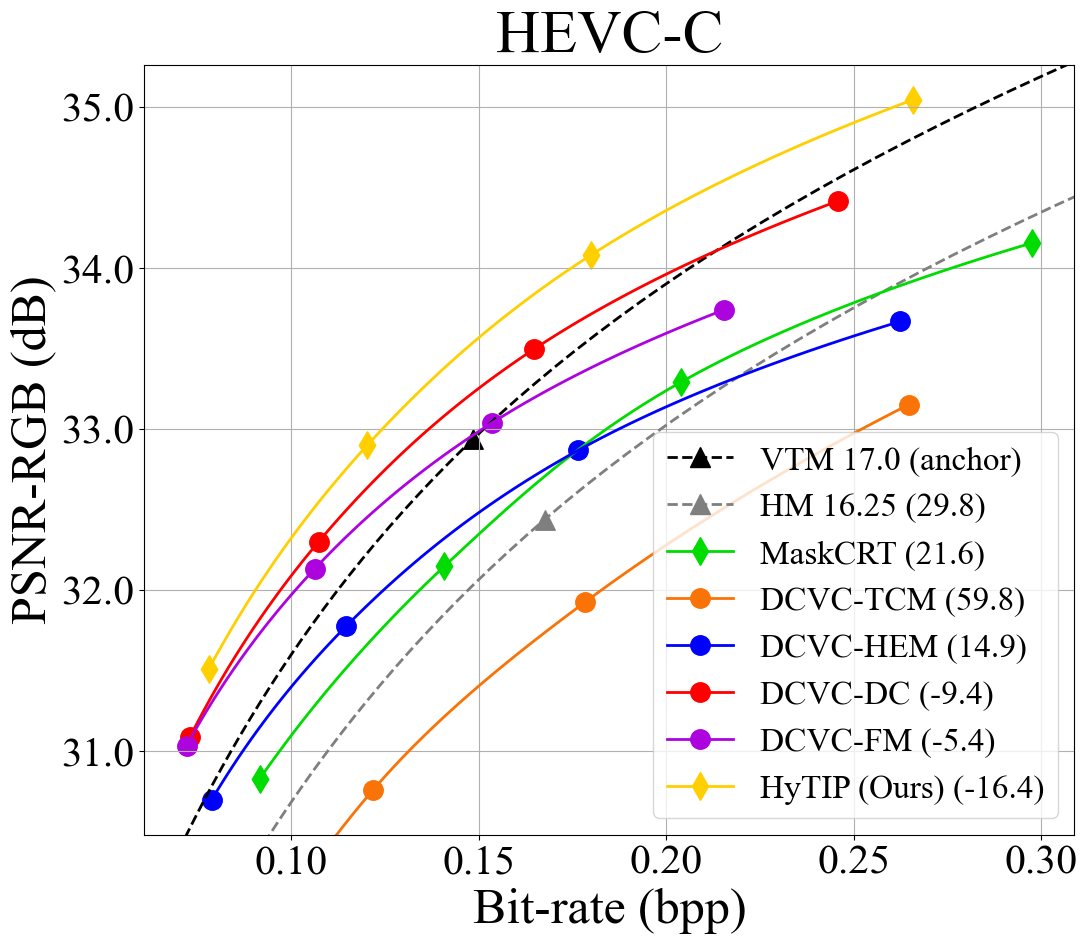}
    \end{subfigure}
    \begin{subfigure}{0.37\linewidth}
        \centering
        \includegraphics[width=\linewidth]{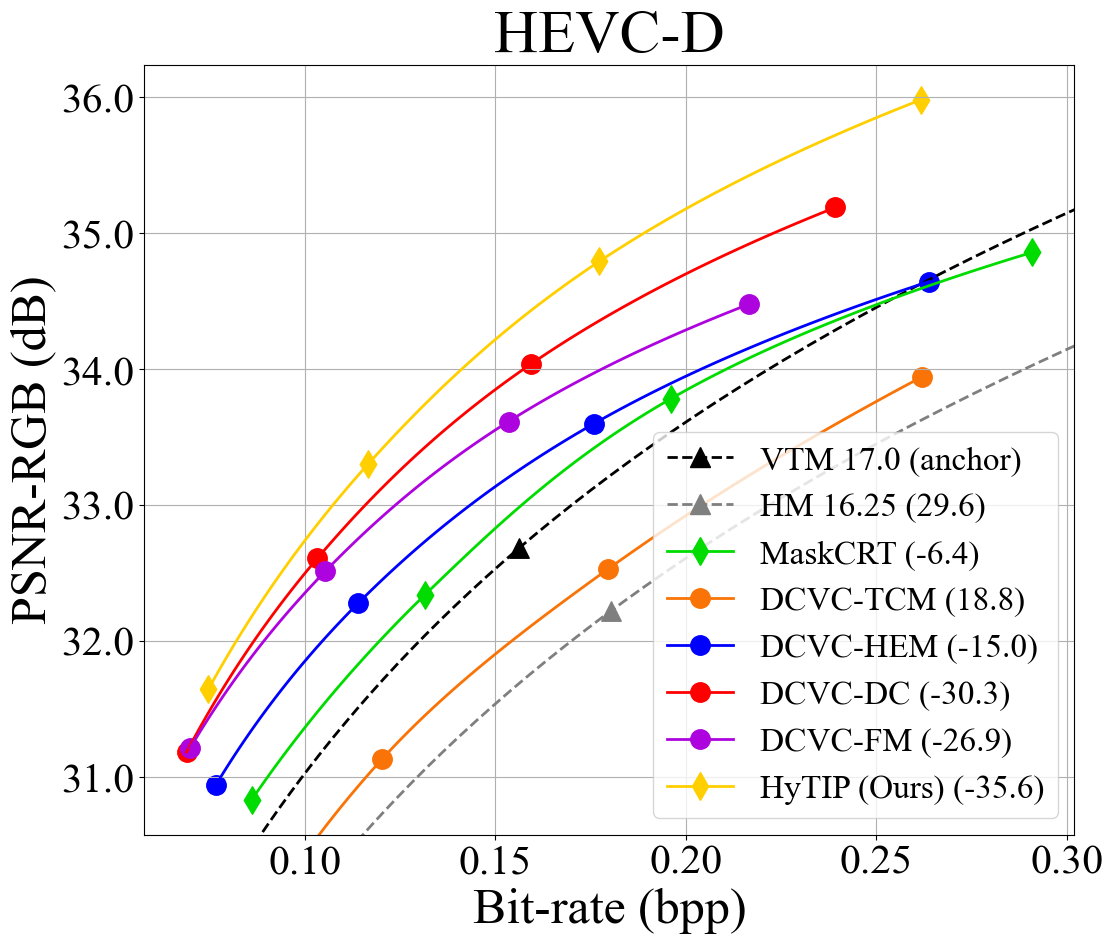}
    \end{subfigure}
    \begin{subfigure}{0.37\linewidth}
        \centering
        \includegraphics[width=\linewidth]{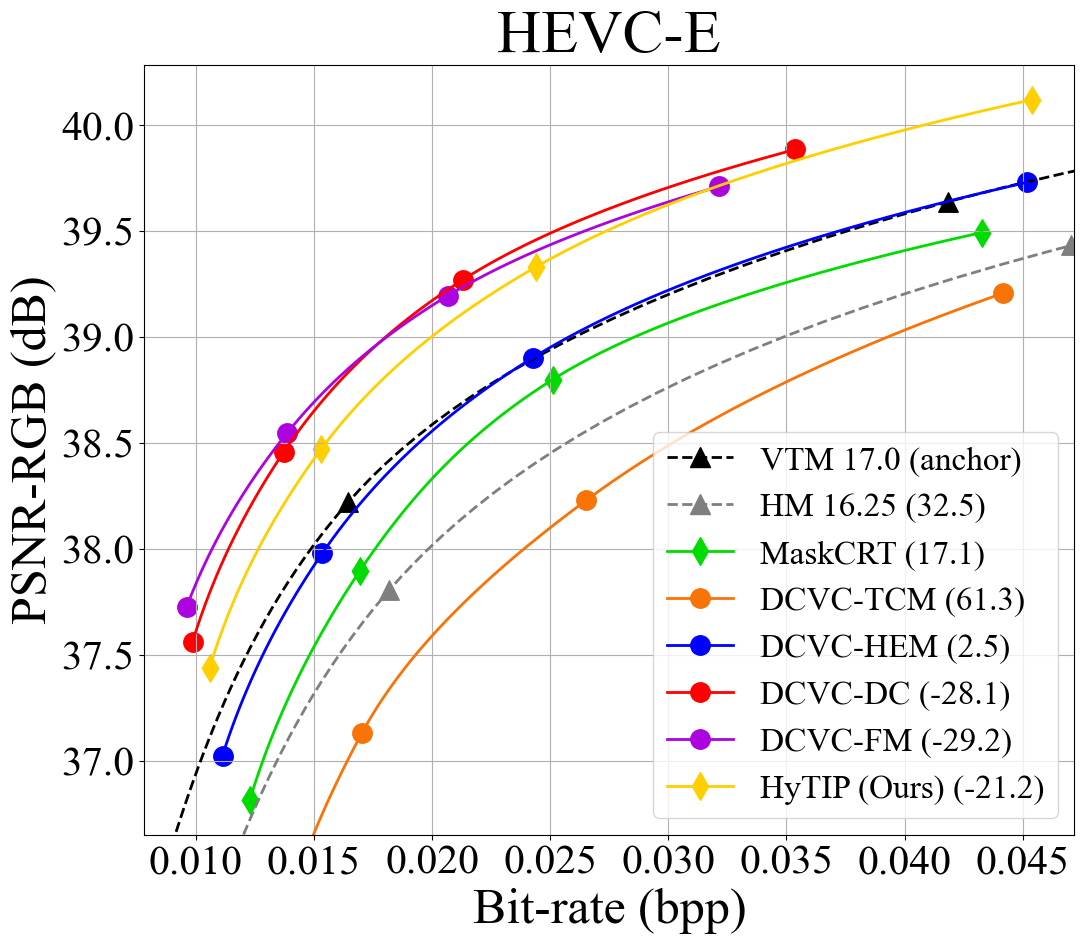}
    \end{subfigure}
    \begin{subfigure}{0.37\linewidth}
        \centering
        \includegraphics[width=\linewidth]{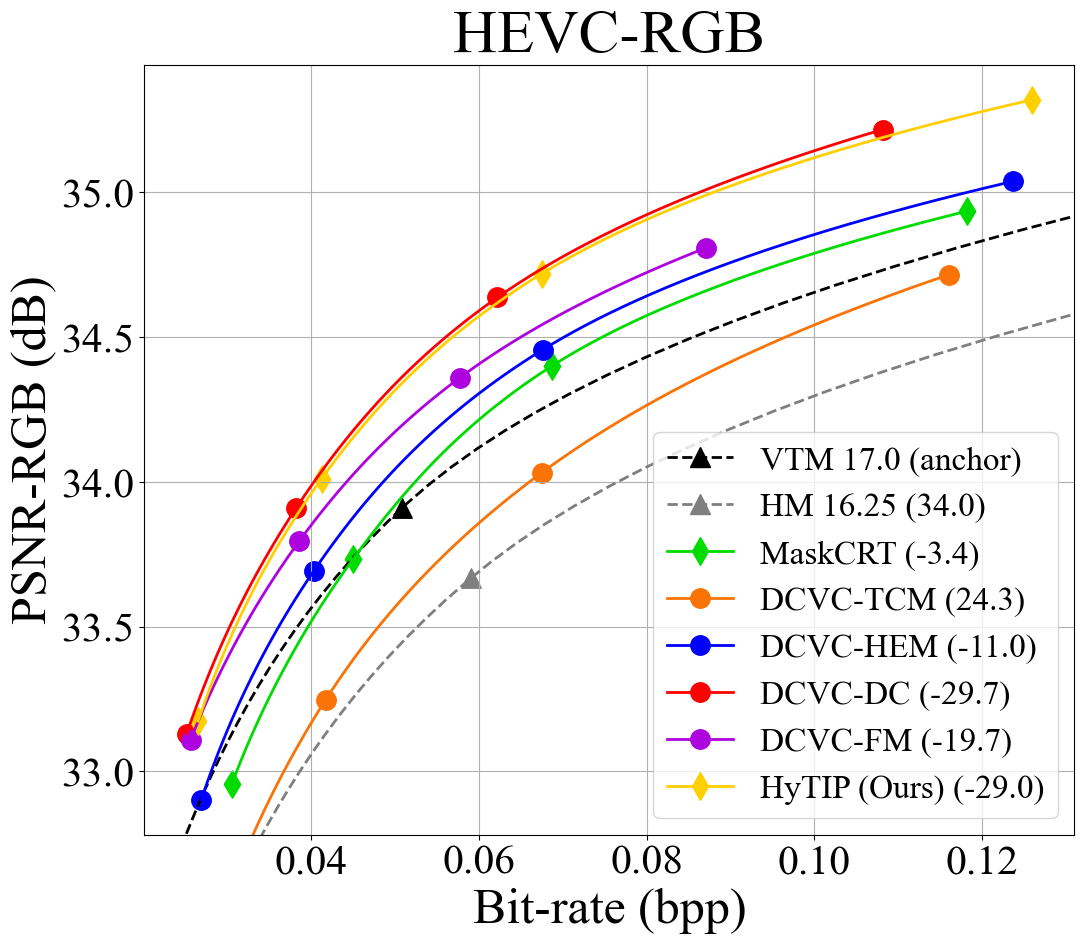}
    \end{subfigure}
    \caption{Rate-distortion comparison with state-of-the-art methods in terms of PSNR-RGB, using BT.601 for color space conversion. The values in parentheses represent BD-rates, with VTM 17.0 serving as the anchor.}
    \label{fig:PSNR_I32_BT601}
    \end{center}
\end{figure*}

\begin{figure*}[tbp]
    \vspace{-2em}
    \begin{center}
    \begin{subfigure}{0.37\linewidth}
        \centering
        \includegraphics[width=\linewidth]{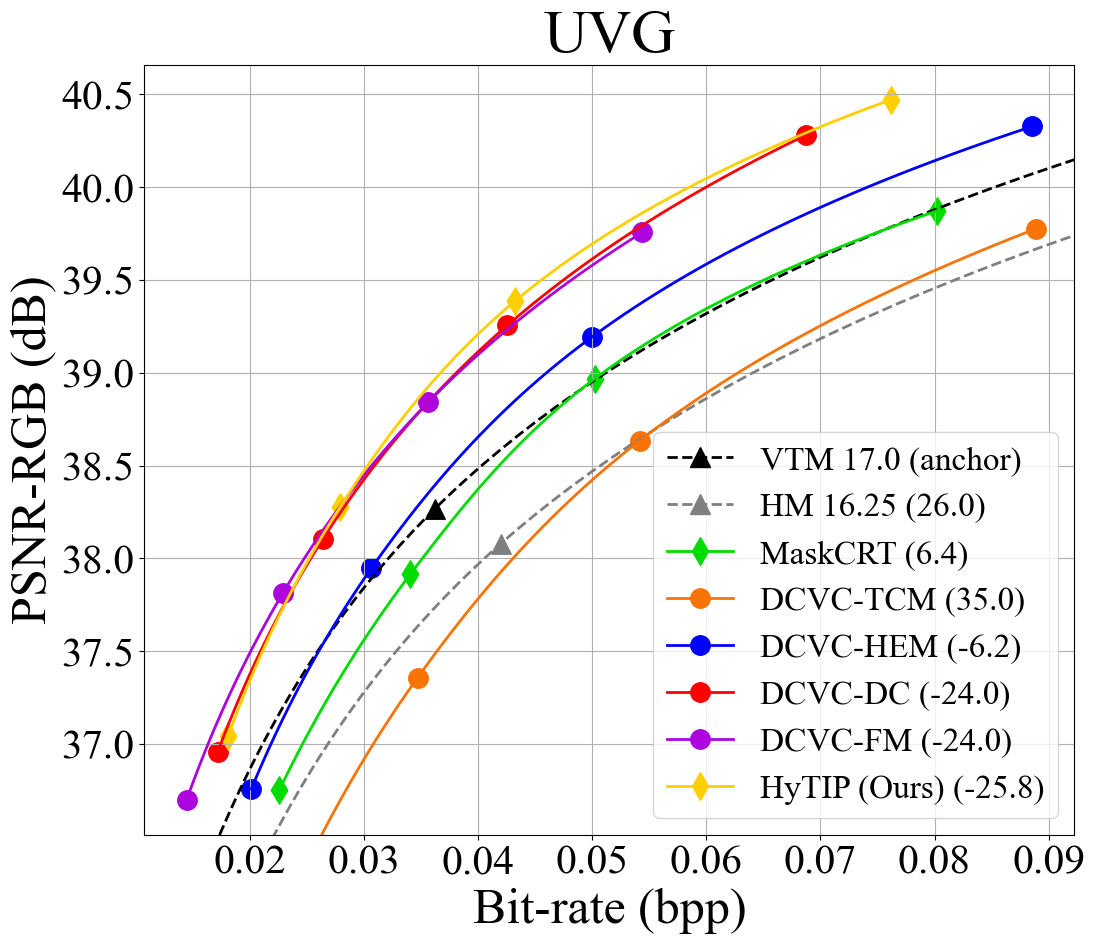}
    \end{subfigure}
    \begin{subfigure}{0.37\linewidth}
        \centering
        \includegraphics[width=\linewidth]{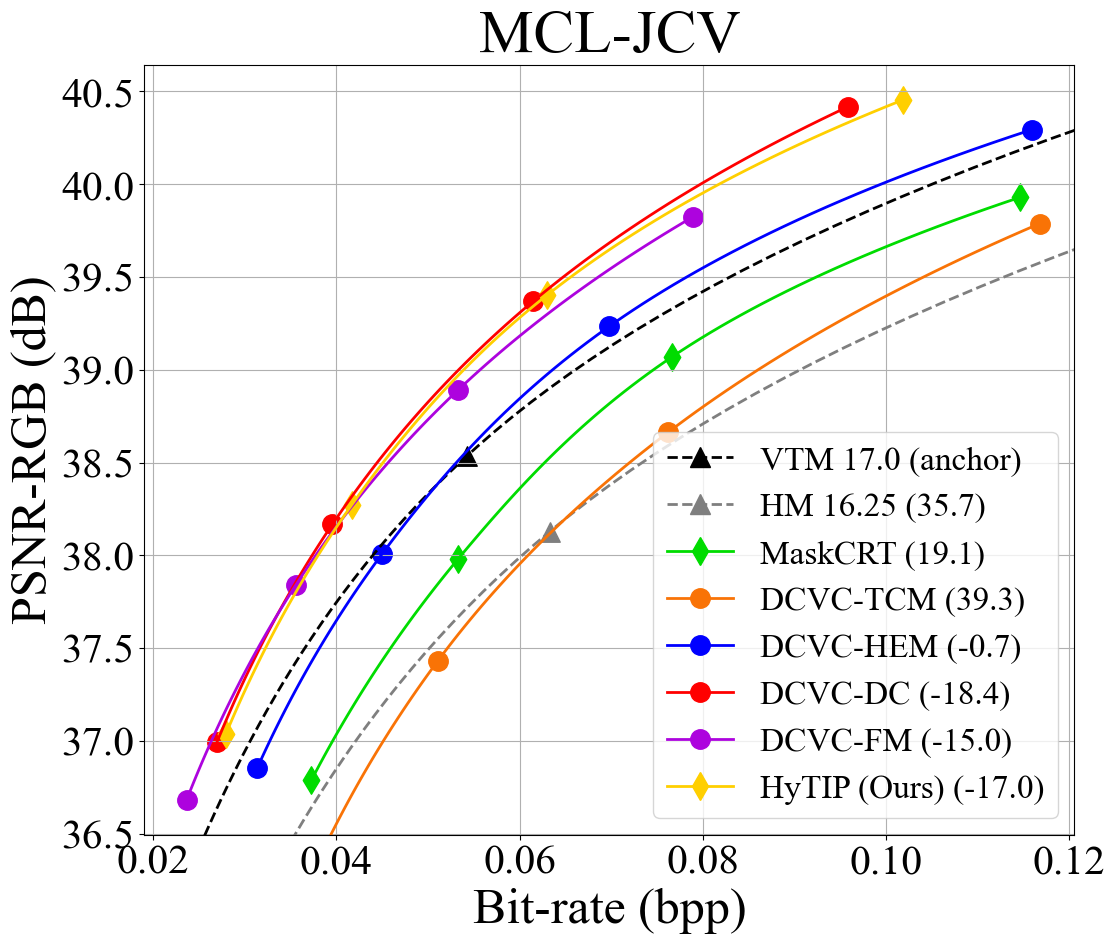}
    \end{subfigure}
    \begin{subfigure}{0.37\linewidth}
        \centering
        \includegraphics[width=\linewidth]{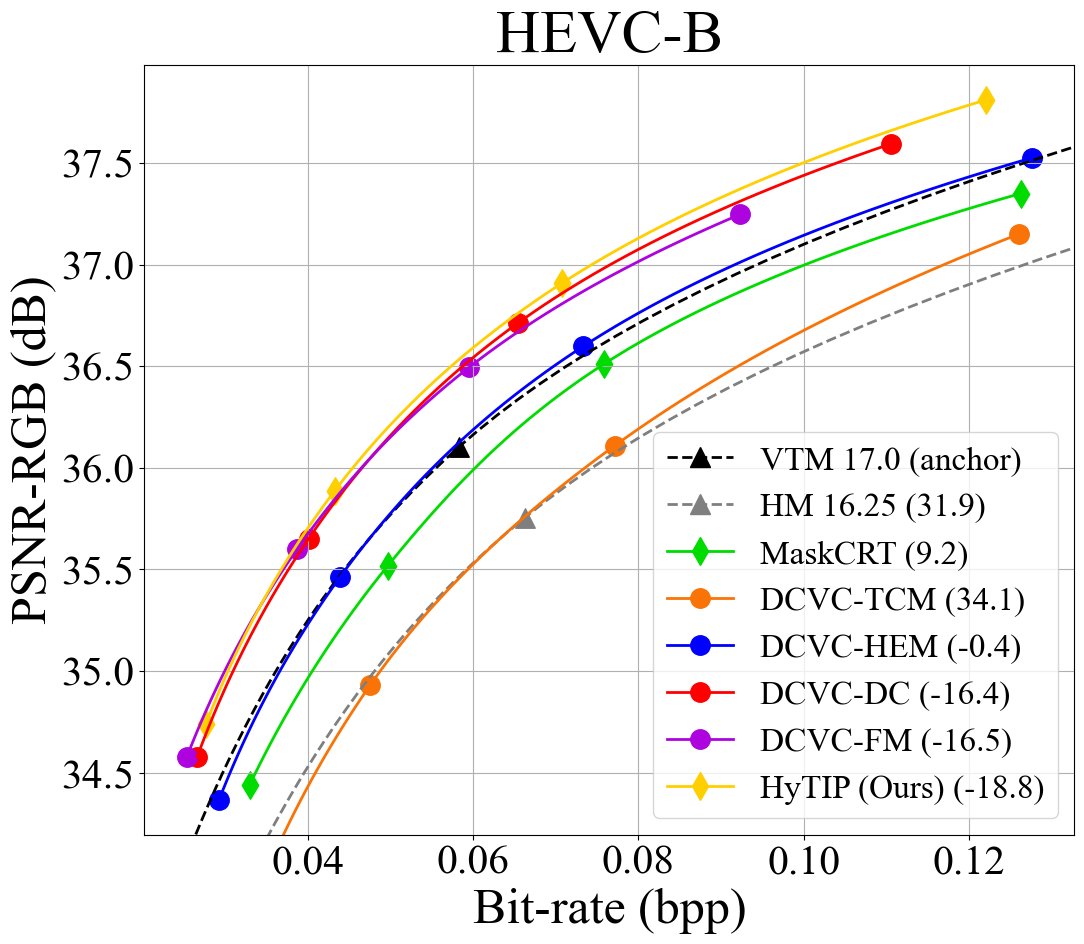}
    \end{subfigure}
    \begin{subfigure}{0.37\linewidth}
        \centering
        \includegraphics[width=\linewidth]{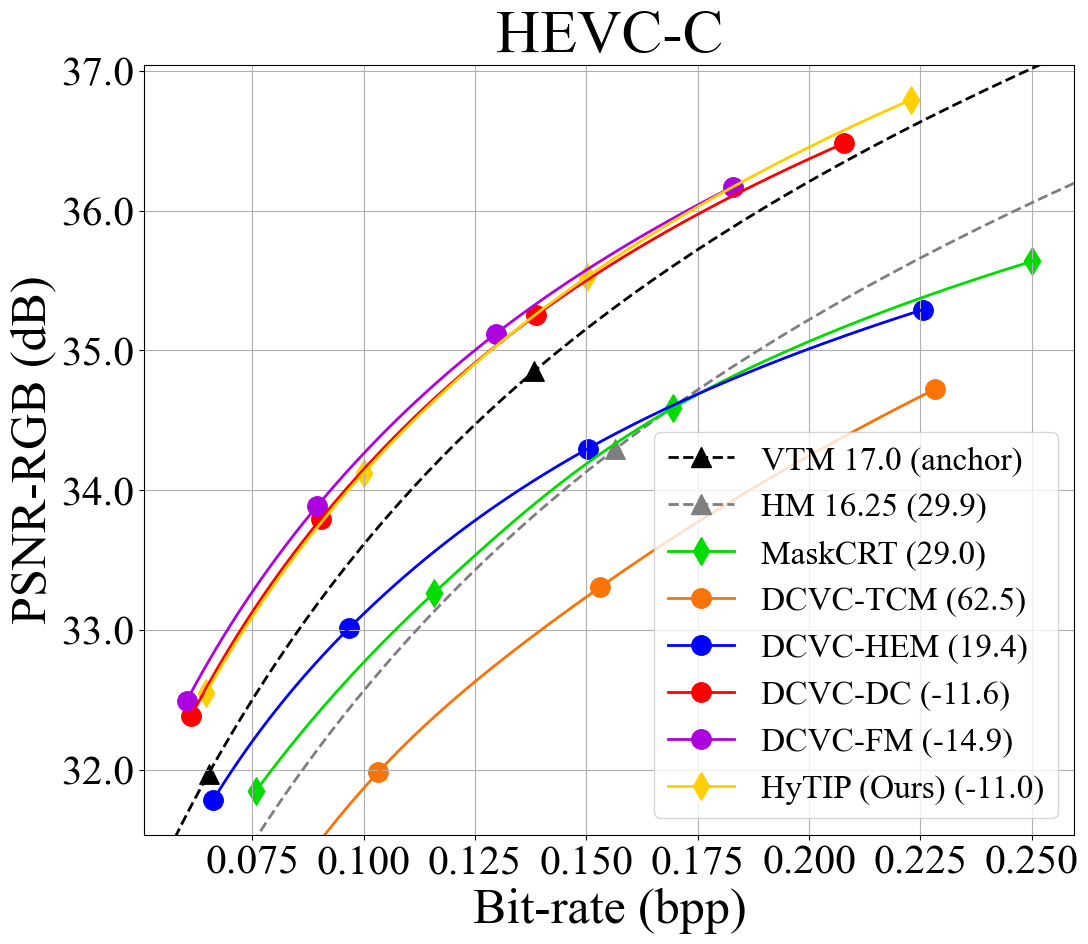}
    \end{subfigure}
    \begin{subfigure}{0.37\linewidth}
        \centering
        \includegraphics[width=\linewidth]{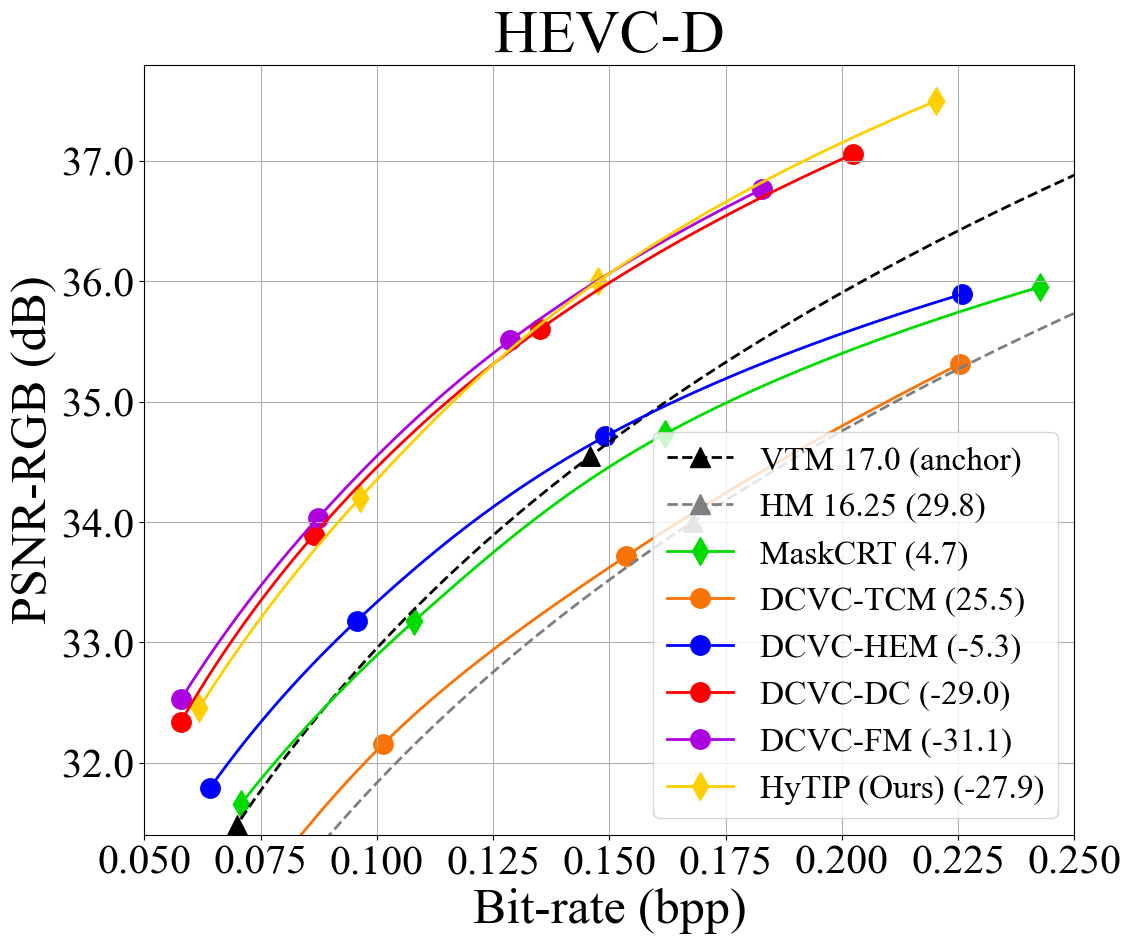}
    \end{subfigure}
    \begin{subfigure}{0.37\linewidth}
        \centering
        \includegraphics[width=\linewidth]{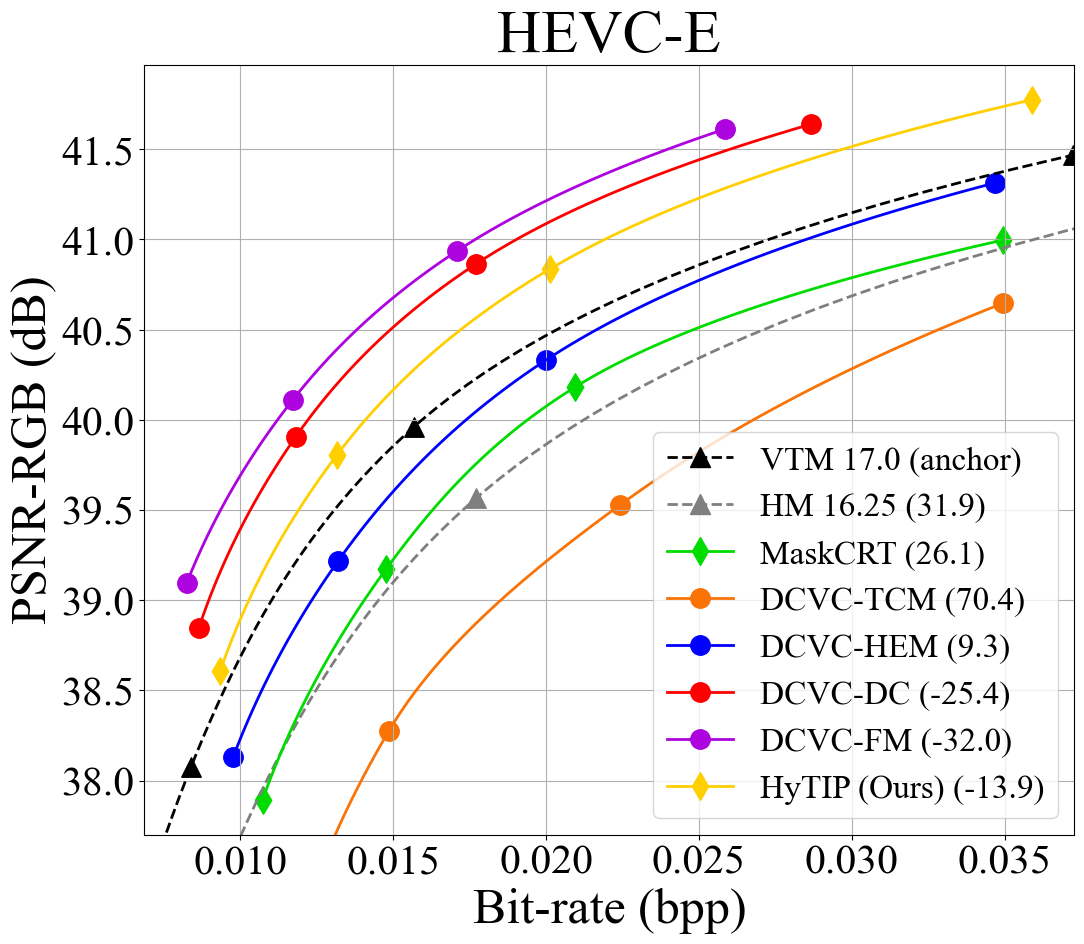}
    \end{subfigure}
    \begin{subfigure}{0.37\linewidth}
        \centering
        \includegraphics[width=\linewidth]{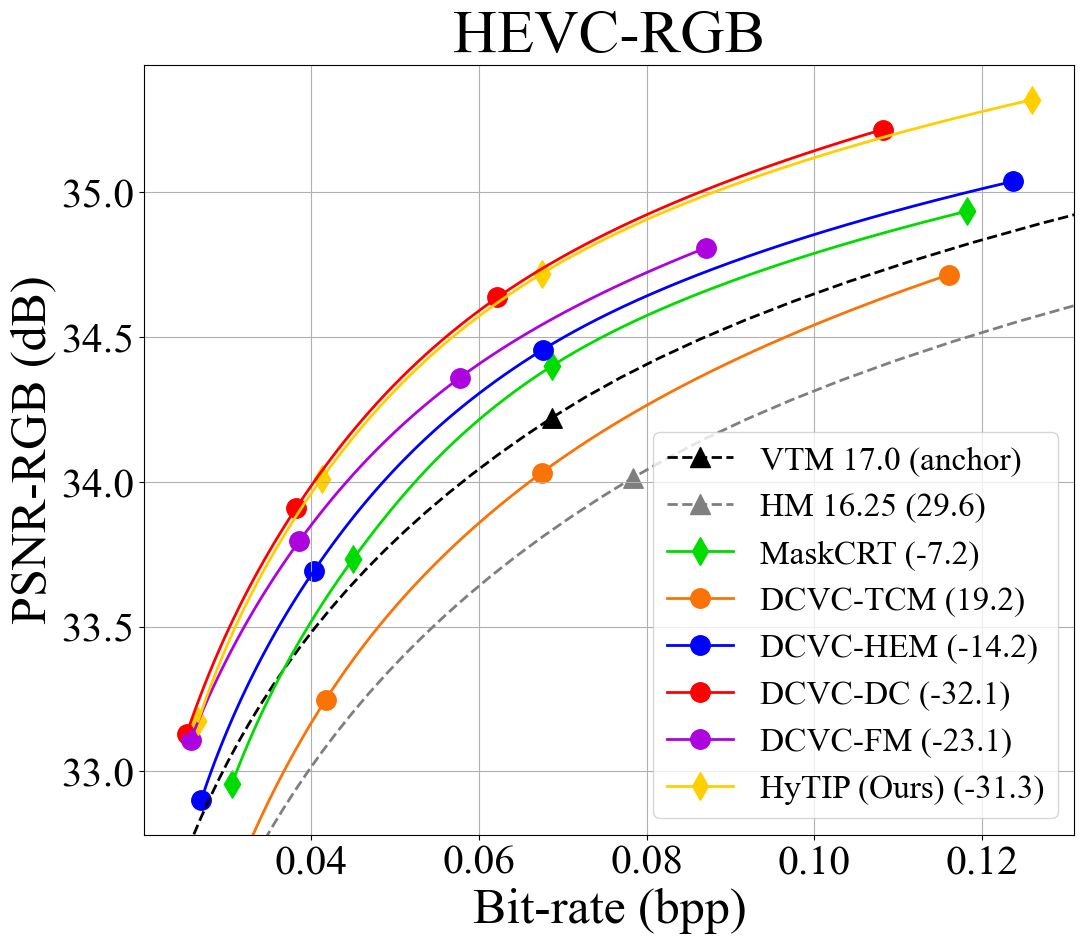}
    \end{subfigure}
    \caption{Rate-distortion comparison with state-of-the-art methods in terms of PSNR-RGB, using BT.709 for color space conversion. The values in parentheses represent BD-rates, with VTM 17.0 serving as the anchor.}
    \label{fig:PSNR_I32_BT709}
    \end{center}
\end{figure*}

\begin{figure*}[tbp]
    \vspace{-2em}
    \begin{center}
    \begin{subfigure}{0.37\linewidth}
        \centering
        \includegraphics[width=\linewidth]{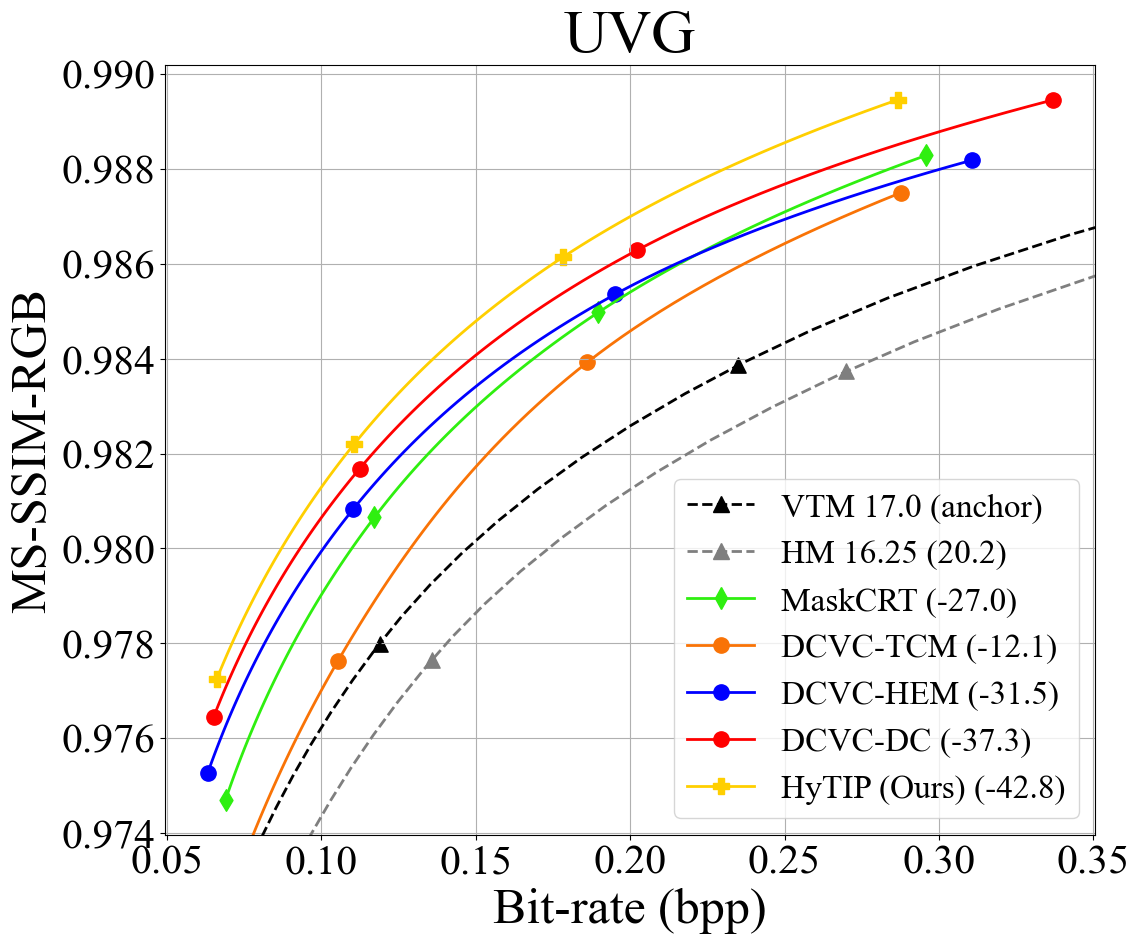}
    \end{subfigure}
    \begin{subfigure}{0.37\linewidth}
        \centering
        \includegraphics[width=\linewidth]{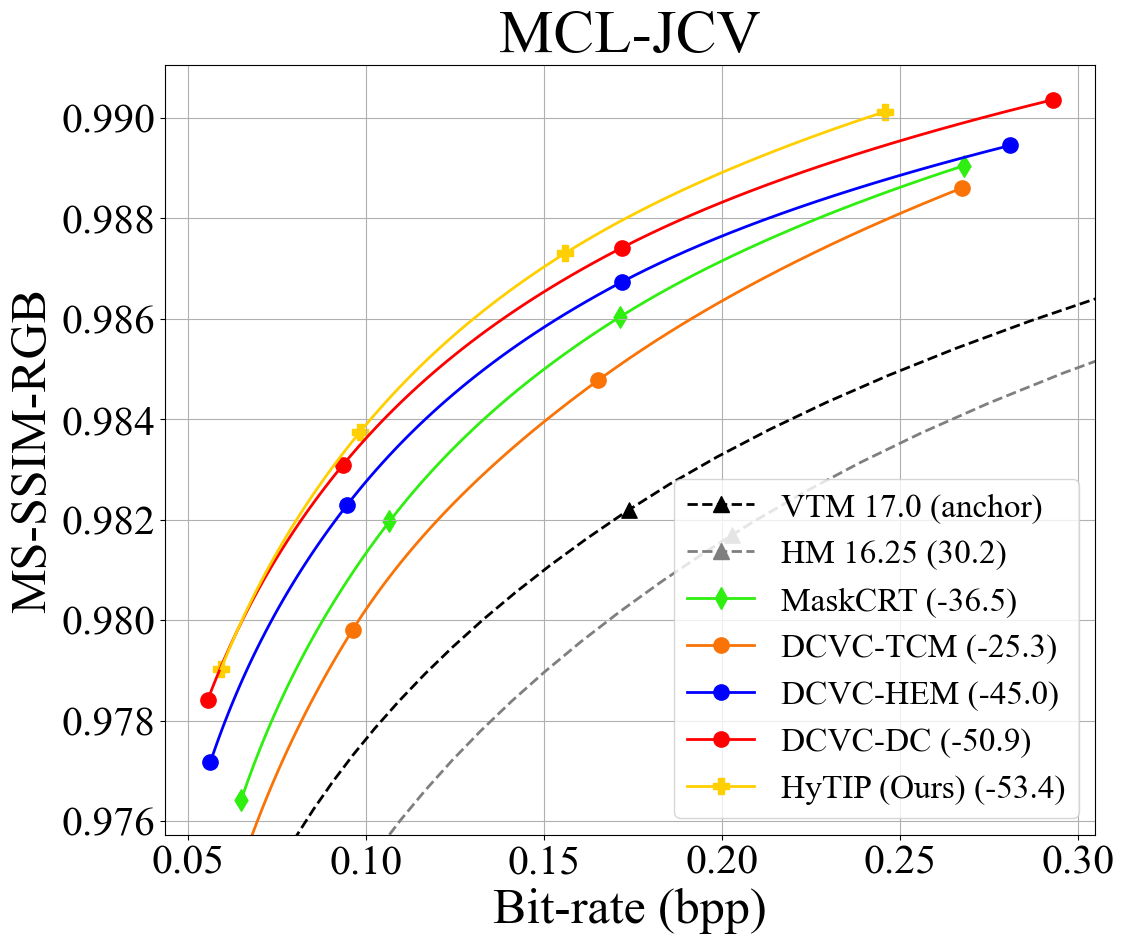}
    \end{subfigure}
    \begin{subfigure}{0.37\linewidth}
        \centering
        \includegraphics[width=\linewidth]{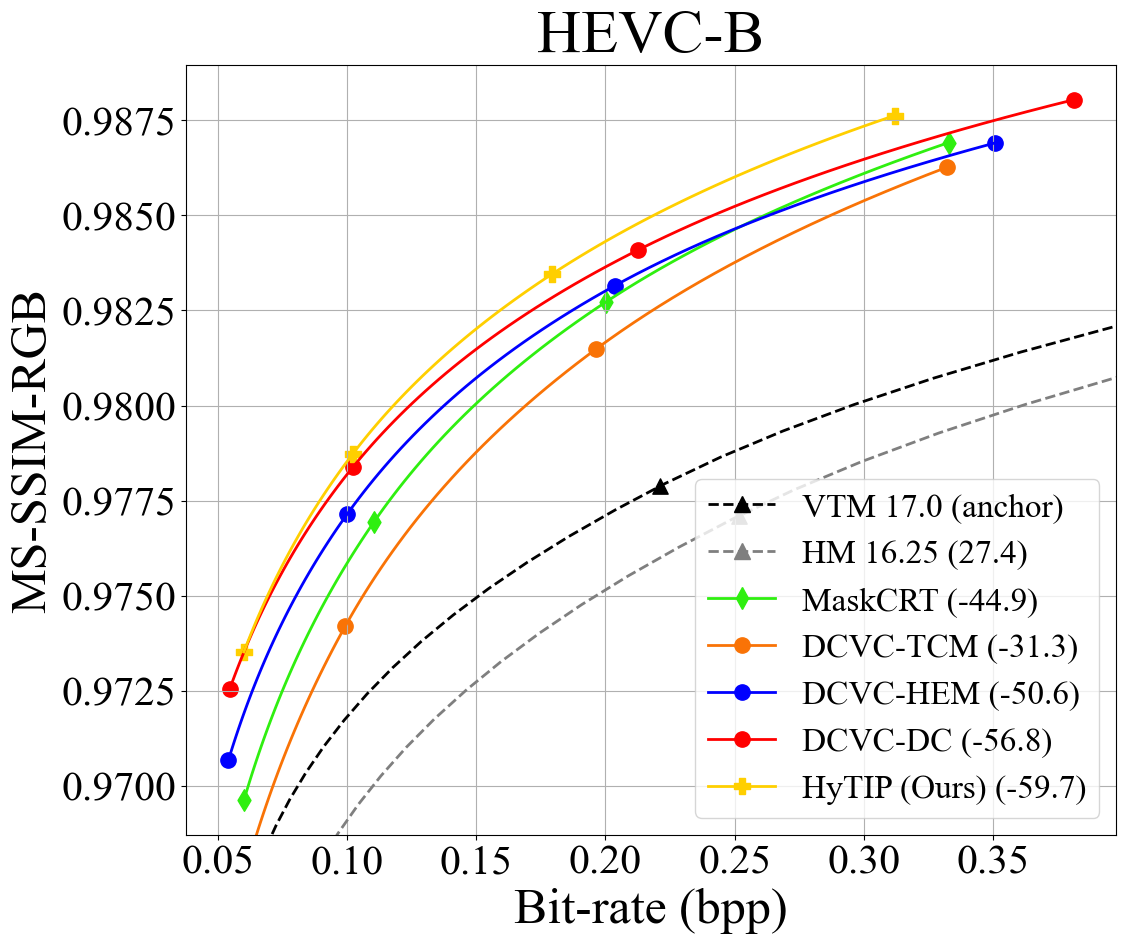}
    \end{subfigure}
    \begin{subfigure}{0.37\linewidth}
        \centering
        \includegraphics[width=\linewidth]{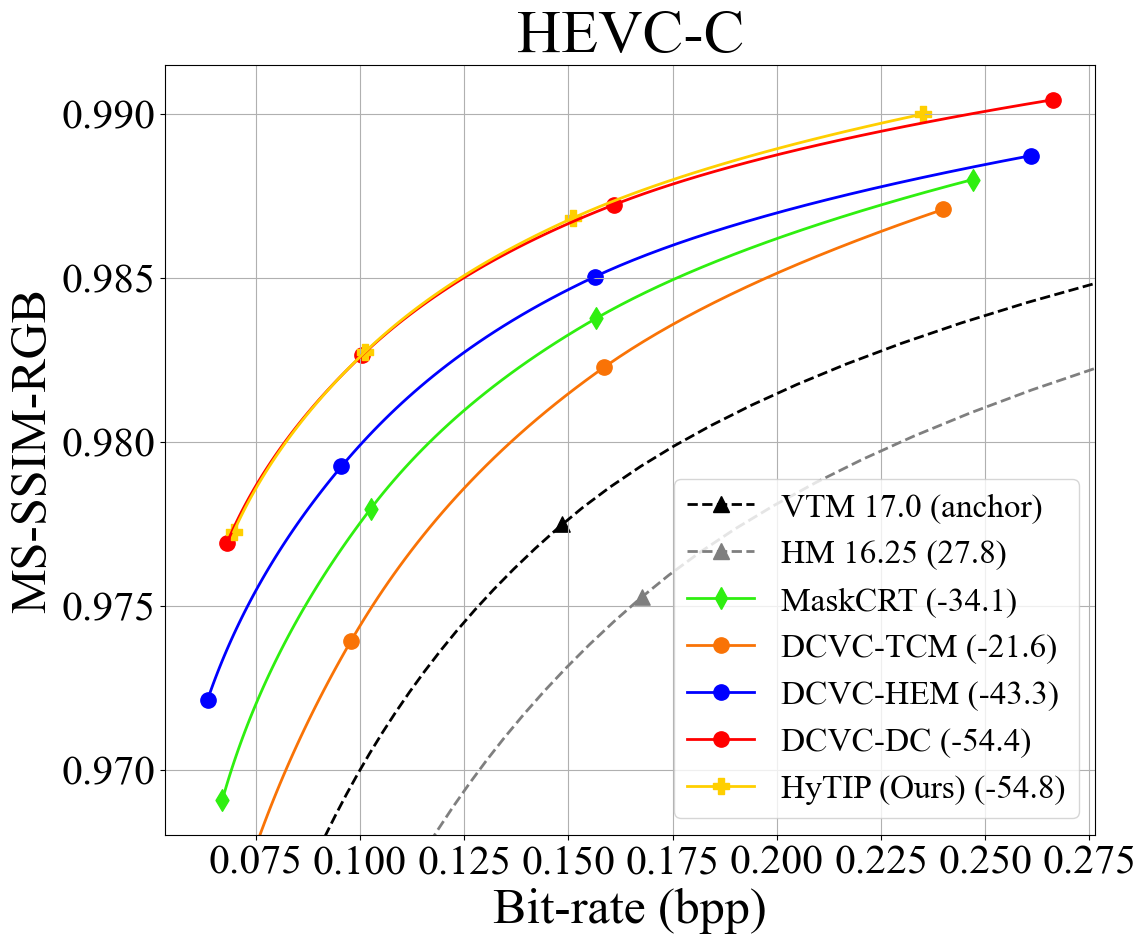}
    \end{subfigure}
    \begin{subfigure}{0.37\linewidth}
        \centering
        \includegraphics[width=\linewidth]{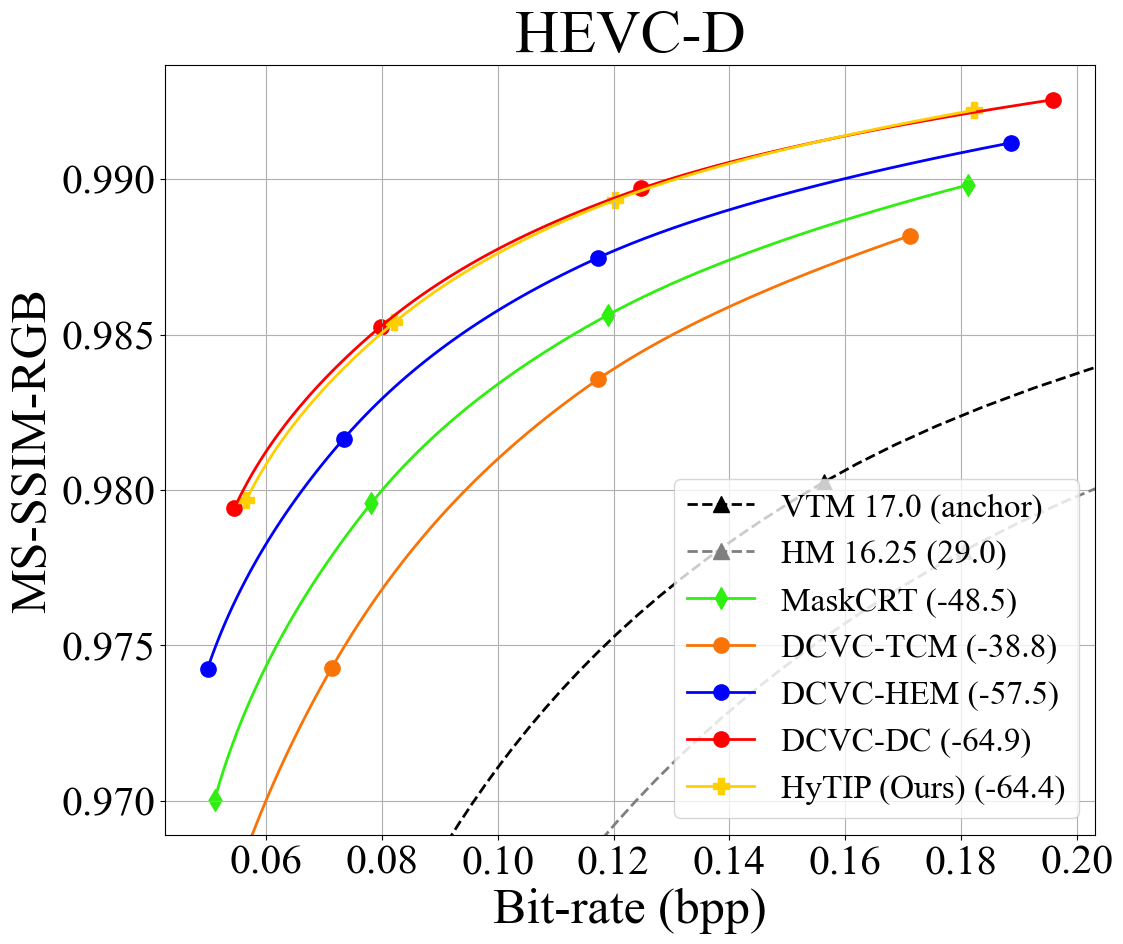}
    \end{subfigure}
    \begin{subfigure}{0.37\linewidth}
        \centering
        \includegraphics[width=\linewidth]{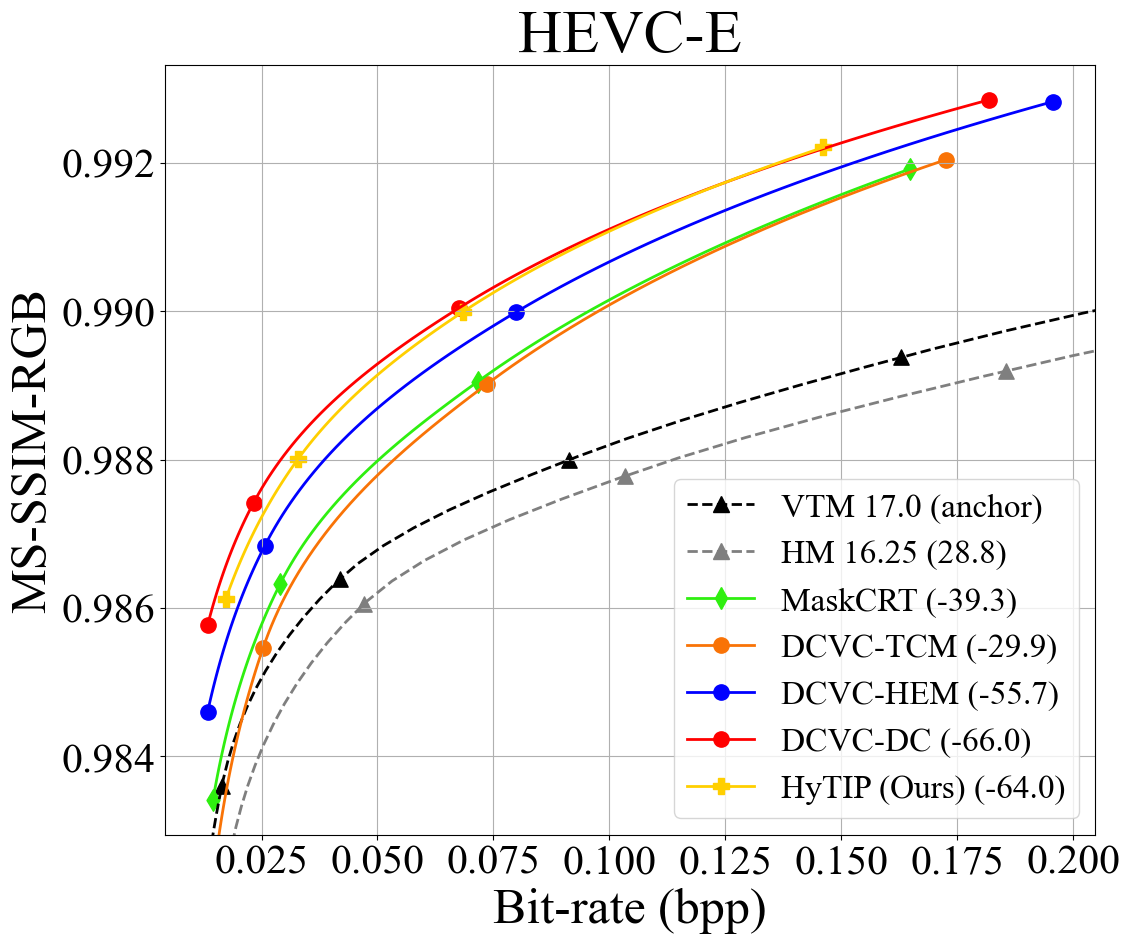}
    \end{subfigure}
    \begin{subfigure}{0.37\linewidth}
        \centering
        \includegraphics[width=\linewidth]{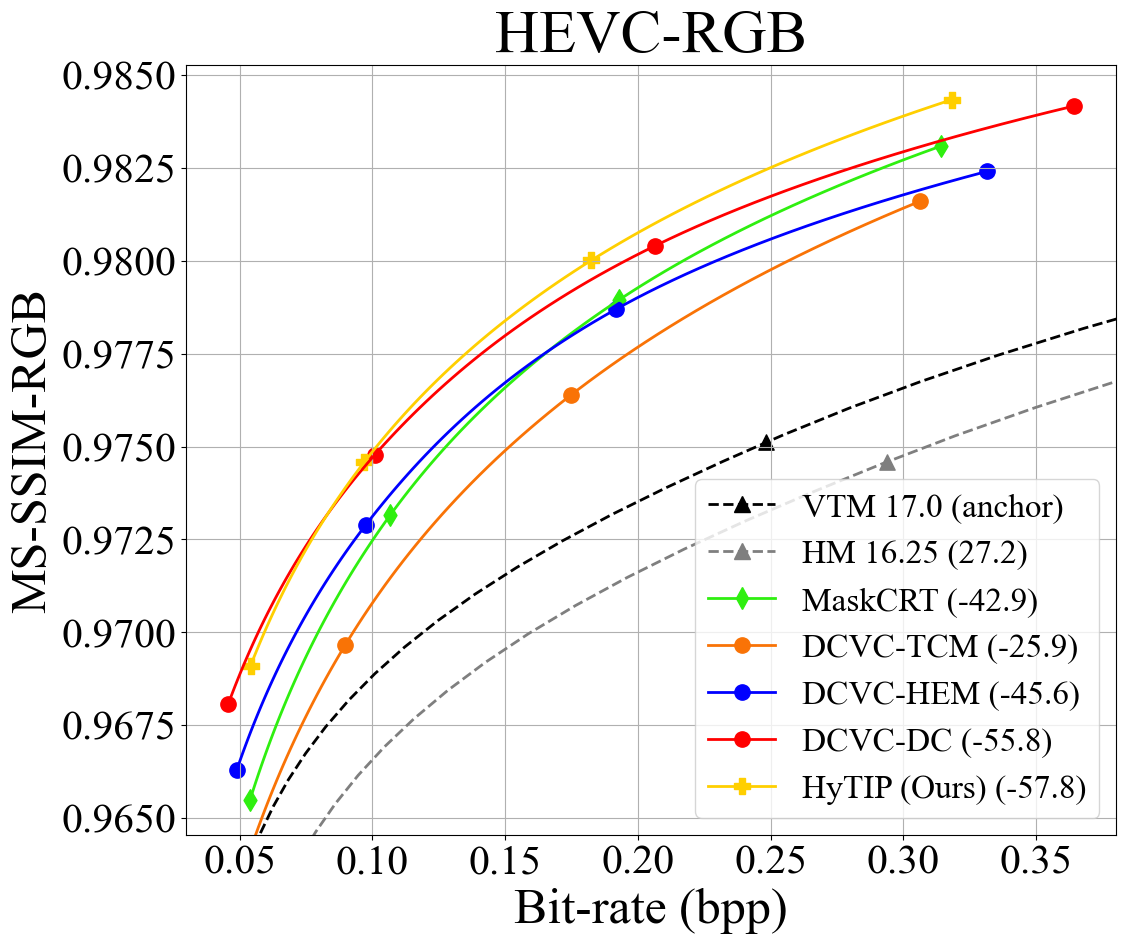}
    \end{subfigure}
    \caption{Rate-distortion comparison with state-of-the-art methods in terms of MS-SSIM-RGB, using BT.601 for color space conversion. The values in parentheses represent BD-rates, with VTM 17.0 serving as the anchor.}
    \label{fig:MS-SSIM_I32_BT601}
    \end{center}
\end{figure*}

\begin{figure*}[tbp]
    \vspace{-2em}
    \begin{center}
    \begin{subfigure}{0.37\linewidth}
        \centering
        \includegraphics[width=\linewidth]{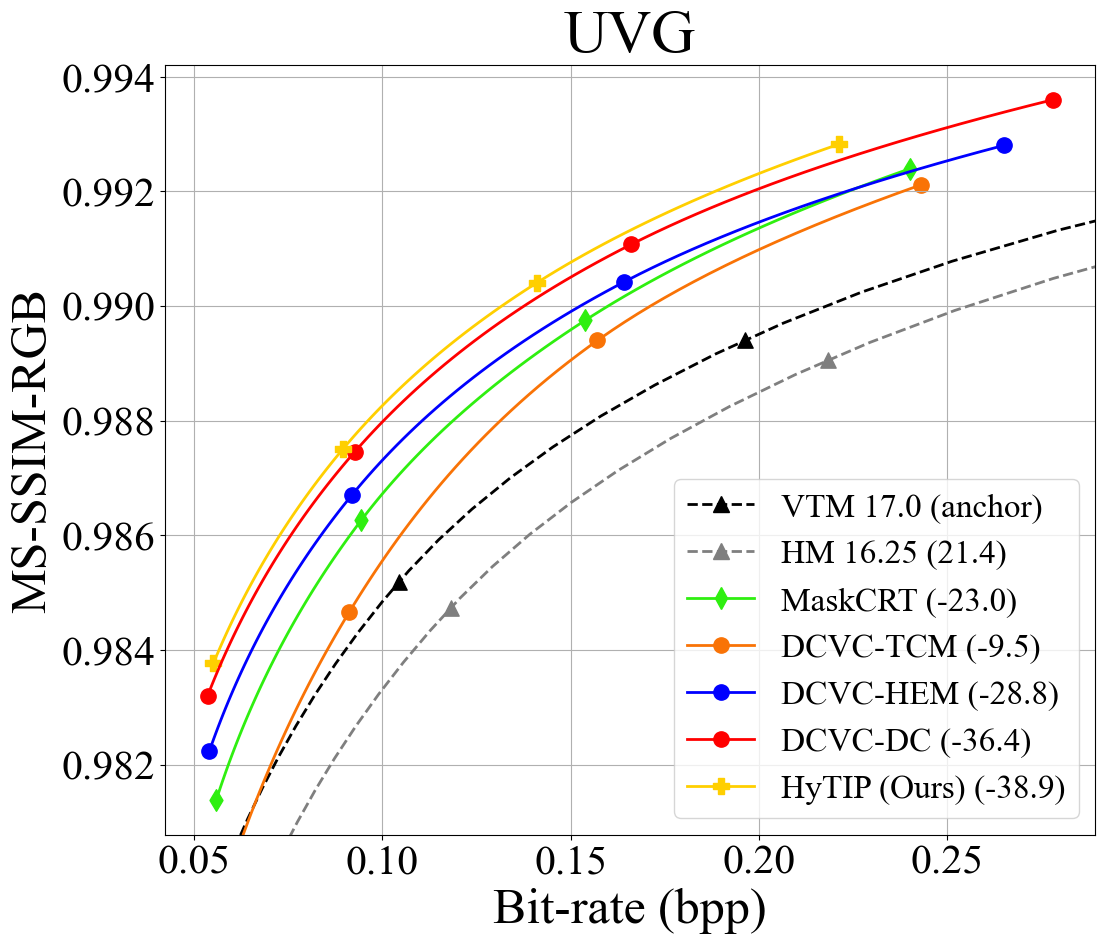}
    \end{subfigure}
    \begin{subfigure}{0.37\linewidth}
        \centering
        \includegraphics[width=\linewidth]{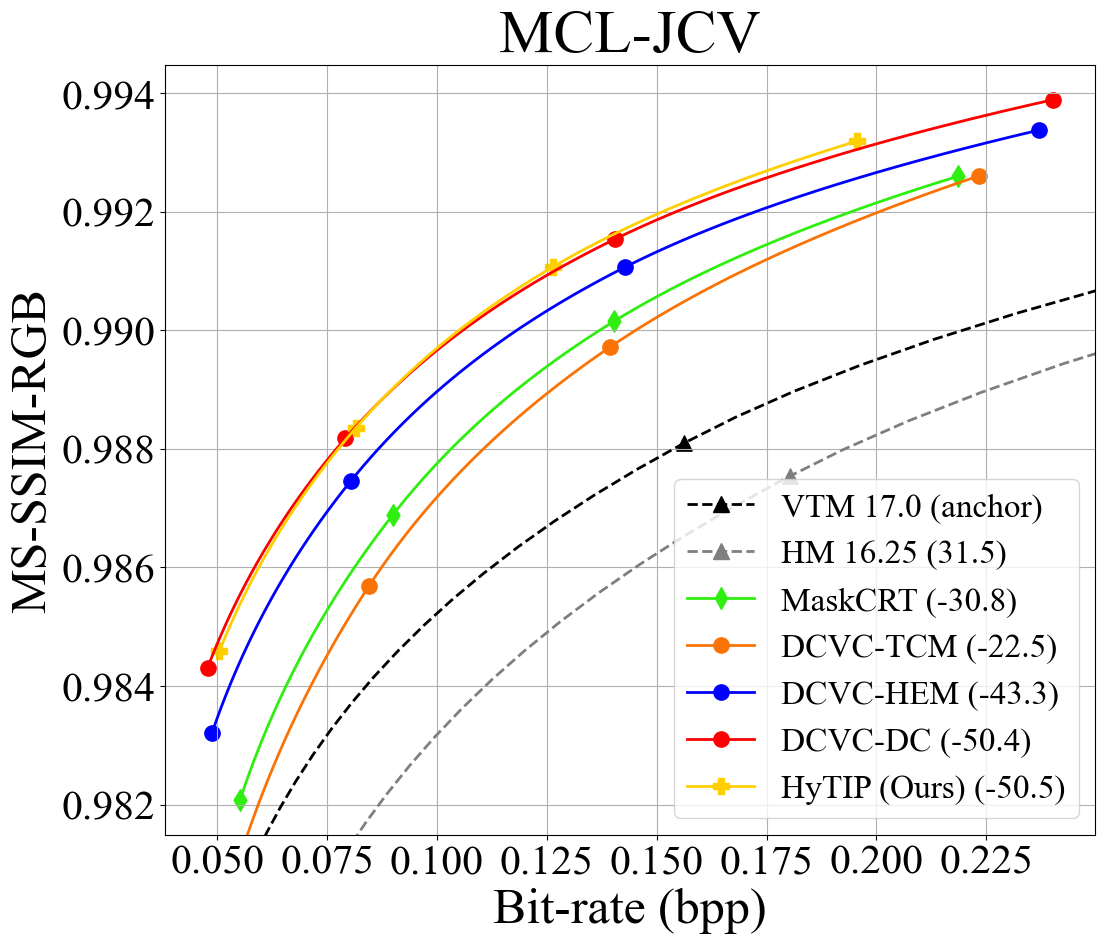}
    \end{subfigure}
    \begin{subfigure}{0.37\linewidth}
        \centering
        \includegraphics[width=\linewidth]{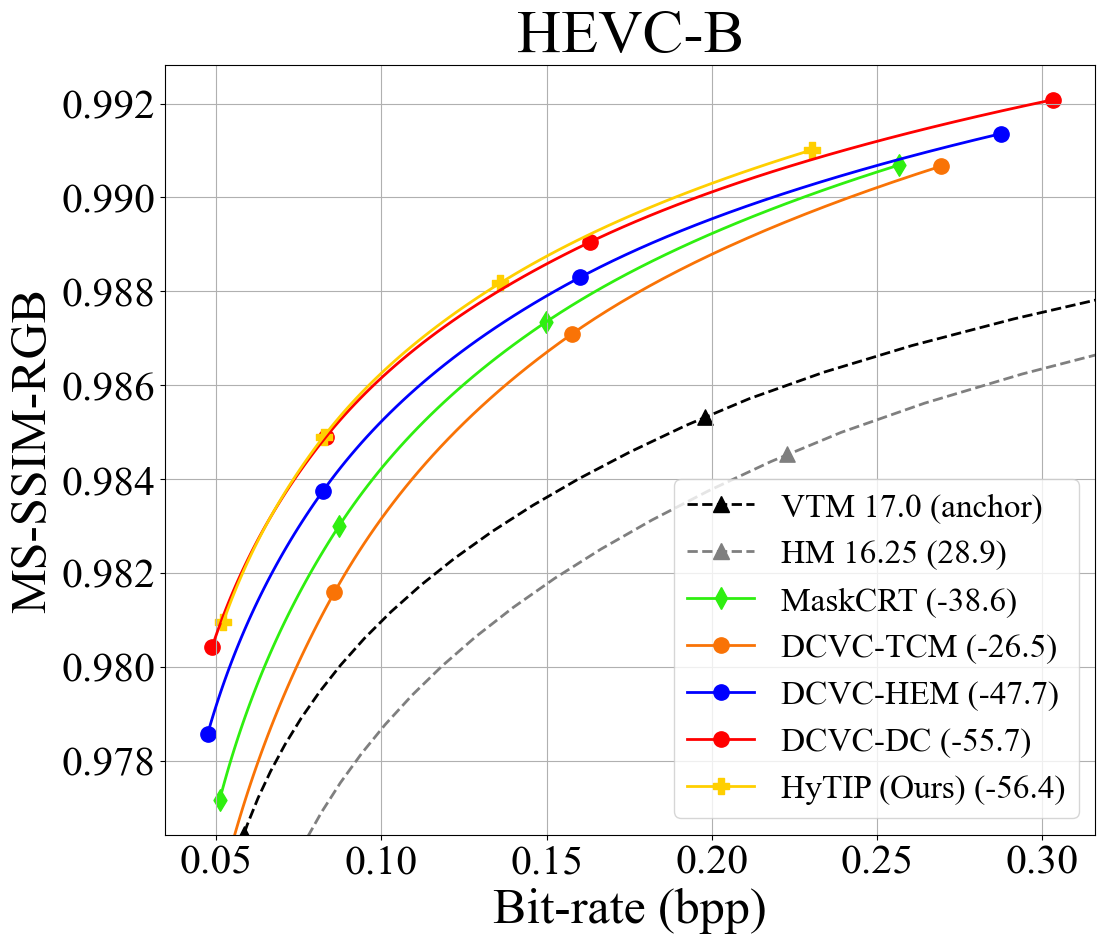}
    \end{subfigure}
    \begin{subfigure}{0.37\linewidth}
        \centering
        \includegraphics[width=\linewidth]{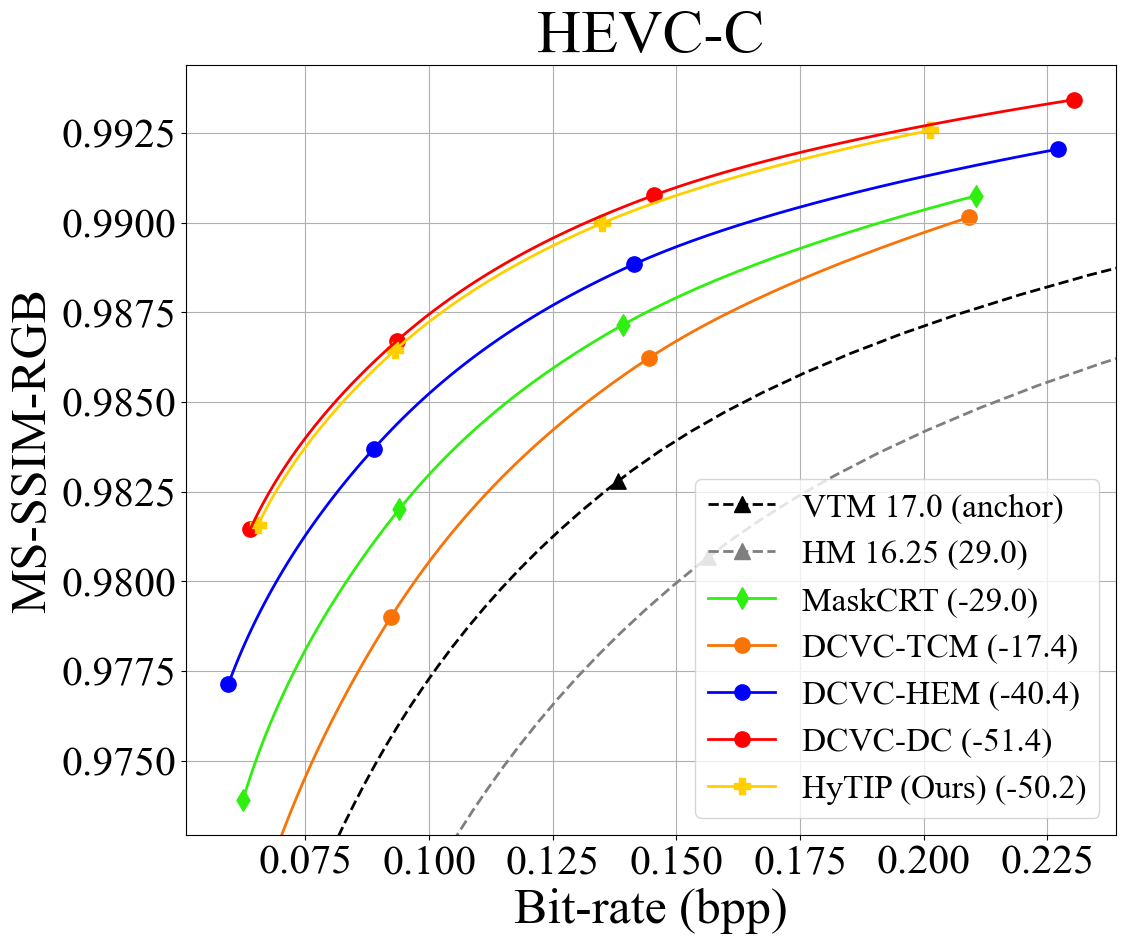}
    \end{subfigure}
    \begin{subfigure}{0.37\linewidth}
        \centering
        \includegraphics[width=\linewidth]{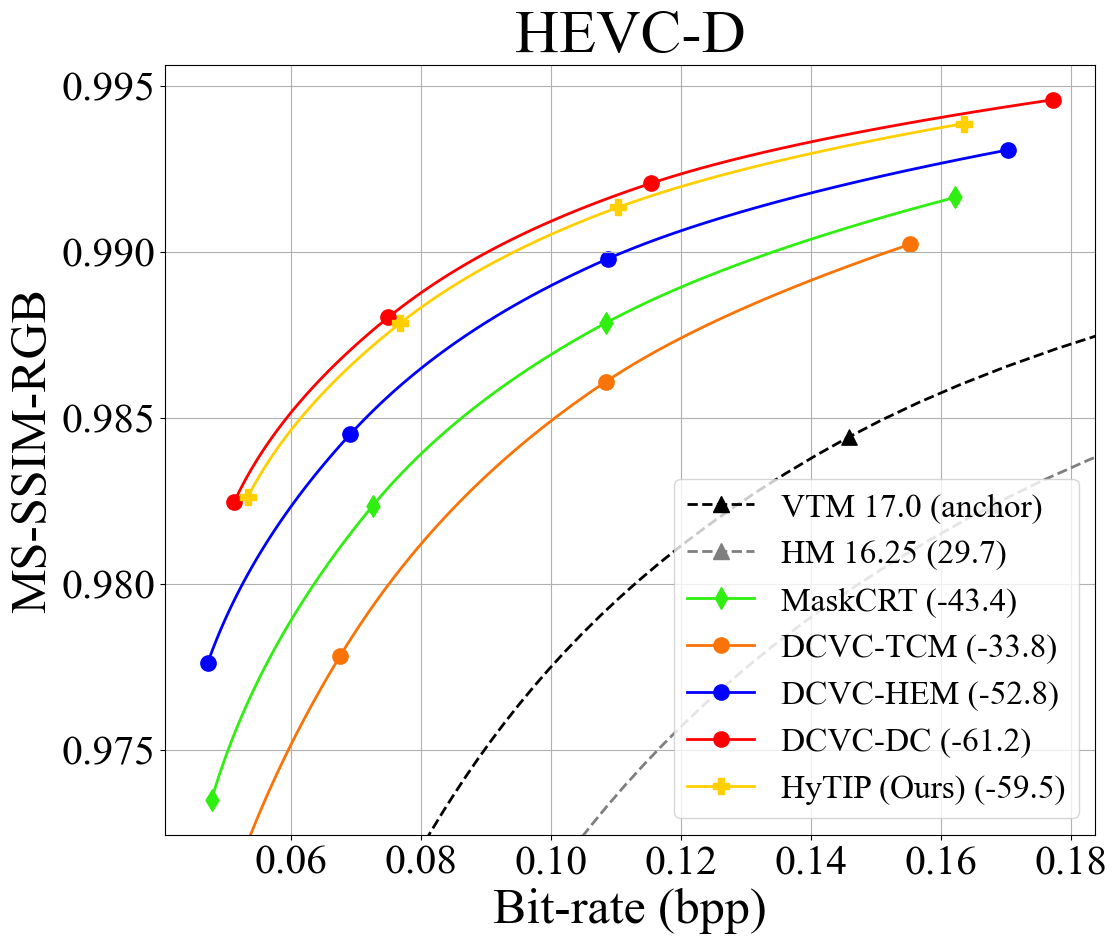}
    \end{subfigure}
    \begin{subfigure}{0.37\linewidth}
        \centering
        \includegraphics[width=\linewidth]{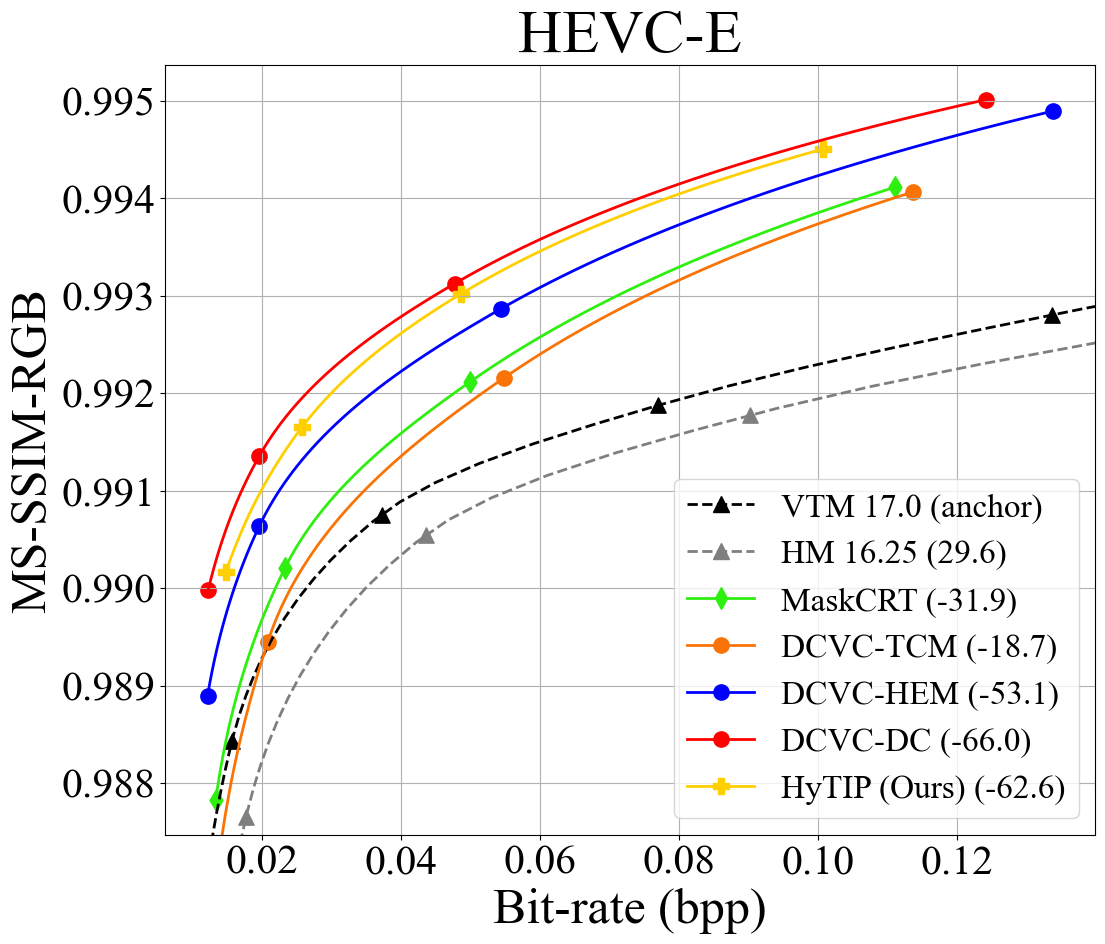}
    \end{subfigure}
    \begin{subfigure}{0.37\linewidth}
        \centering
        \includegraphics[width=\linewidth]{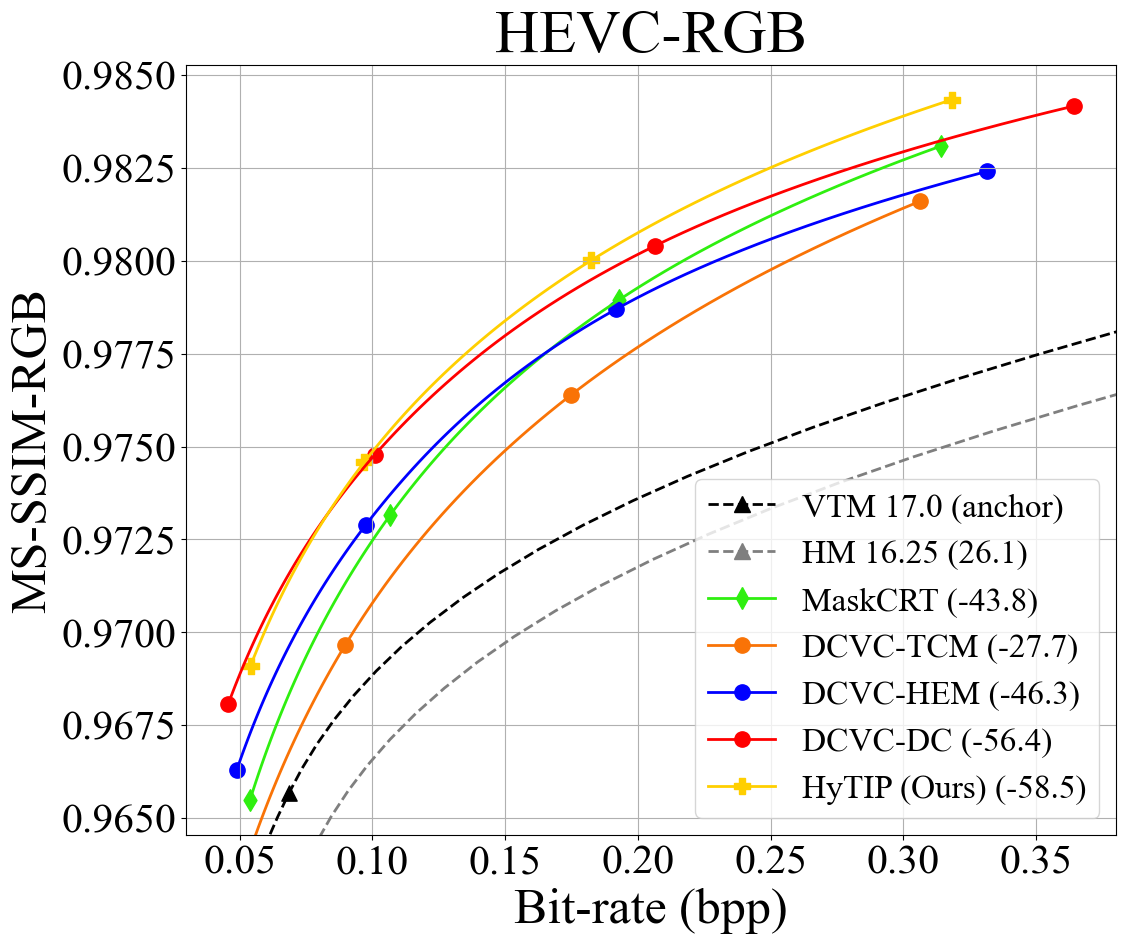}
    \end{subfigure}
    \caption{Rate-distortion comparison with state-of-the-art methods in terms of MS-SSIM-RGB, using BT.709 for color space conversion. The values in parentheses represent BD-rates, with VTM 17.0 serving as the anchor.}
    \label{fig:MS-SSIM_I32_BT709}
    \end{center}
\end{figure*}

Fig.~\ref{fig:PSNR_I32_BT601} and Fig.~\ref{fig:PSNR_I32_BT709} present the rate-distortion comparisons between our method and state-of-the-art approaches in terms of PSNR-RGB, using BT.601 and BT.709 for color space conversion, respectively. Similarly, Fig.~\ref{fig:MS-SSIM_I32_BT601} and Fig.~\ref{fig:MS-SSIM_I32_BT709} show the comparisons in terms of MS-SSIM-RGB under BT.601 and BT.709 color space conversion, respectively. \new{The corresponding BD-rates using BT.709 as the color space conversion are summarized in Table~\ref{table:sota_psnr_bt709_gop32} and Table~\ref{table:sota_msssim_bt709_gop32}.}

\section{Network Architecture Details}
\label{sec:architecture}
\begin{figure*}[t!]
    \centering
    \includegraphics[width=1.\linewidth]{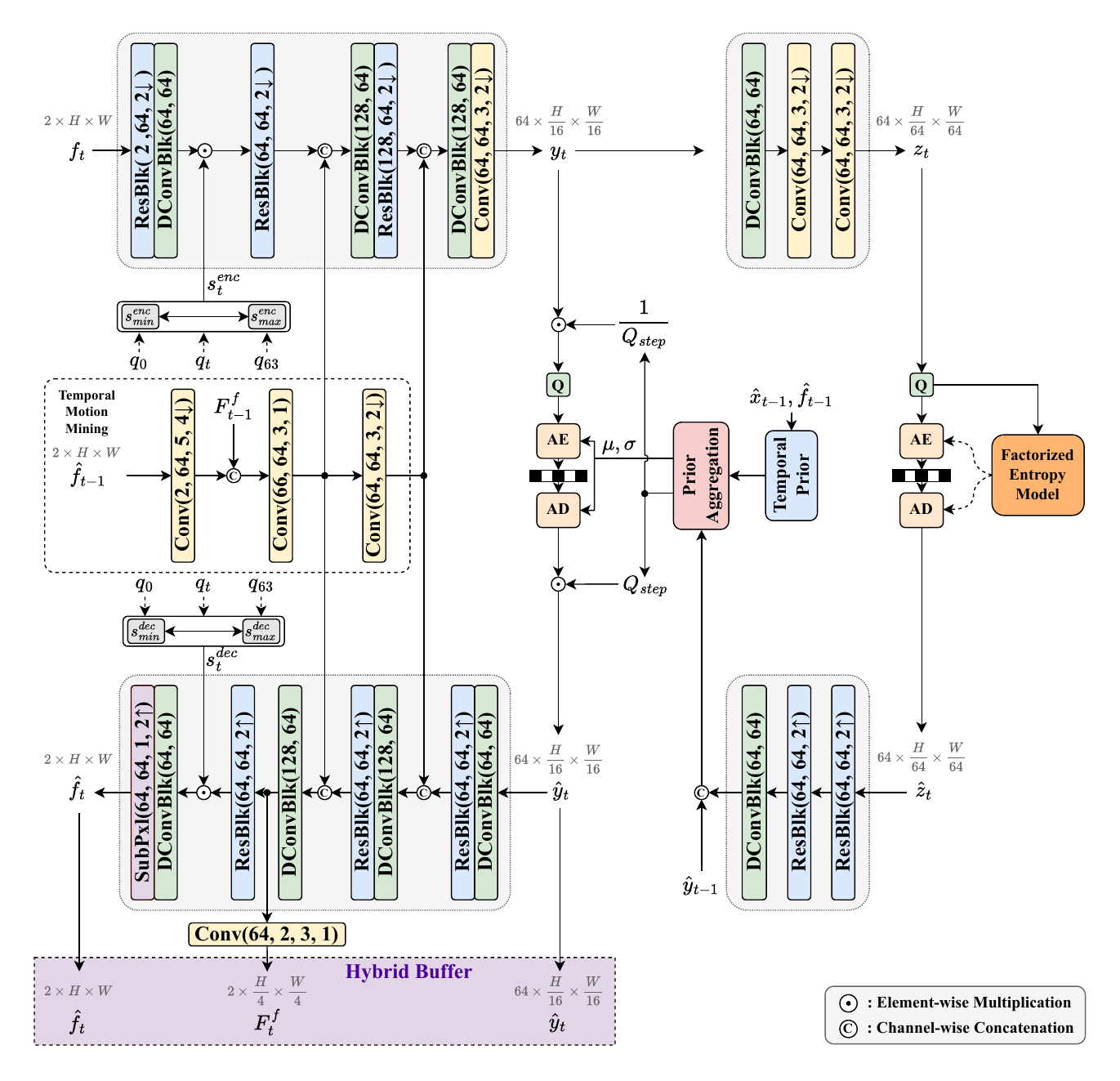}
    \caption{Network architecture detail of our motion codec.}
    \label{fig:motion}
\end{figure*}

\begin{figure*}[t!]
    \centering
    \includegraphics[width=1.\linewidth]{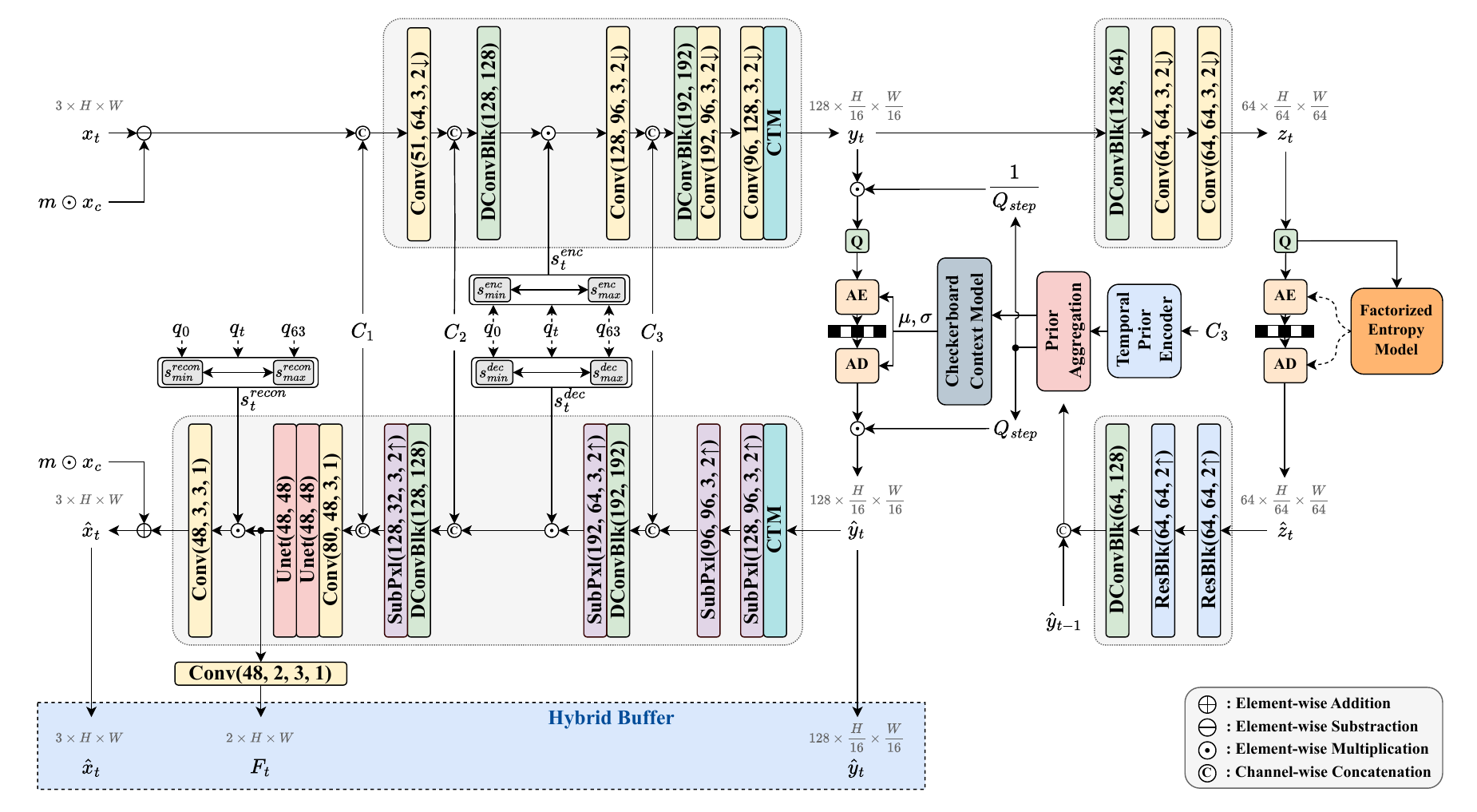}
    \caption{Network architecture detail of our inter-frame codec.}
    \label{fig:inter}
\end{figure*}

\begin{figure*}[t!]
    \centering
    \includegraphics[width=1.\linewidth]{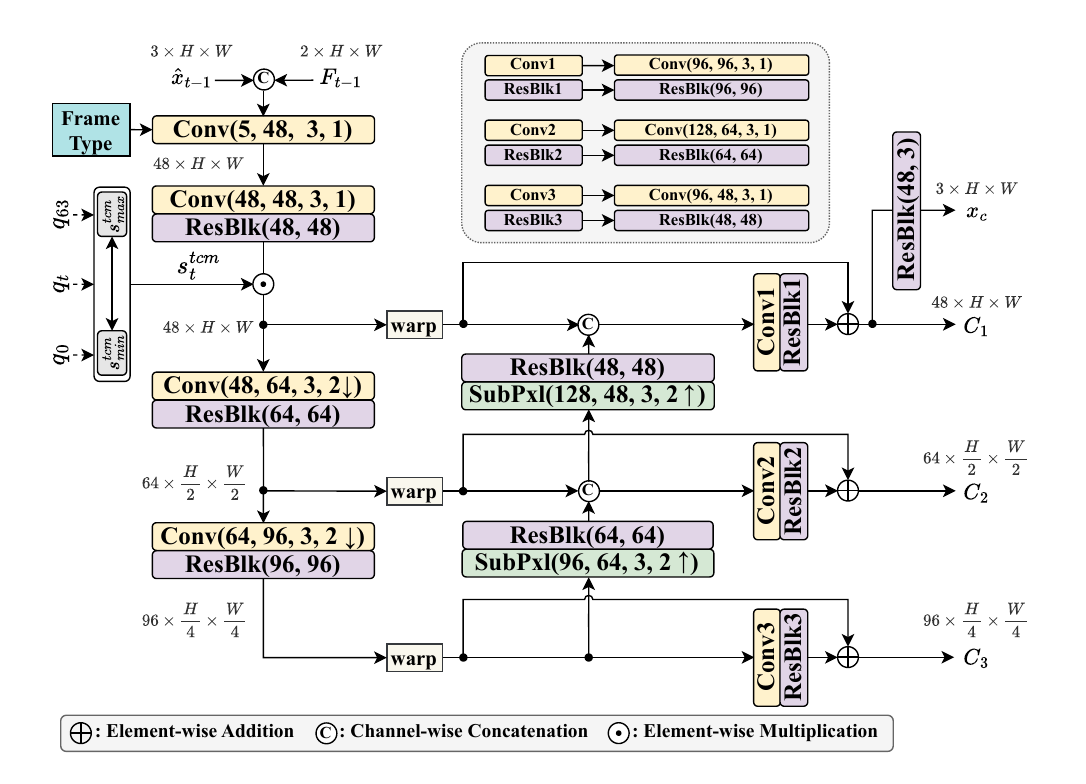}
    \caption{Network architecture detail of our temporal context mining in our inter-frame codec.}
    \label{fig:TCM}
\end{figure*}

Fig.~\ref{fig:motion}, Fig.~\ref{fig:inter}, and Fig.~\ref{fig:TCM} provide the network architecture details in Fig.~\ref{fig:overview} of the main paper. The base motion and inter-frame codec are adapted from \cite{fm} while the channel transform module in the inter-frame codec is from \cite{maskcrt} and the checkerboard context model in the inter-frame codec is from \cite{multistage}.

\section{Training Details}
\label{sec:training_details}
Table~\ref{tab:training} summarizes our HyTIP training procedure, adapted from \cite{maskcrt}. The first six phases follow \cite{maskcrt}, using explicit temporal reference $\hat{x}_{t-1}$ as the temporal reference information only in the inter-frame codec. Subsequently, the implicit related module is incorporated for further training.

\begin{table*}[h]
\fontsize{9pt}{9pt}\selectfont
\caption{BD-rate (\%) comparison of different buffering strategies with different buffer sizes\new{, using BT.601 for color space conversion.} The anchor employs explicit buffering in both motion and inter-frame coding. \new{The values in parentheses, as in Table~\ref{table:buffer_size}, indicate the number of full-resolution feature maps buffered for coding one input flow map or frame (explicit + implicit).} 
}
%
\centering
\label{table:ablation_BT601}
\setlength{\tabcolsep}{5pt}
\begin{tabular}{l|l|rrrrrrr|r}
\toprule
Motion            & Inter         & UVG   & MCL-JCV & HEVC-B & HEVC-C & HEVC-D & HEVC-E & HEVC-RGB & Average \\
\midrule
Explicit (2+0)      & Explicit (3+0)  & 0     & 0       & 0      & 0      & 0      & 0      & 0        & 0       \\
\midrule
Implicit (0+4)      & Explicit (3+0)  & -11.5 & -9.5    & -12.6  & -18.6  & -12.6  & -11.9  & -8.8     & -12.2   \\
Implicit (0+2.1875) & Explicit (3+0)  &  -7.0 & -3.6    &  -8.7  & -12.0  &  -8.4  &  -9.8  & -7.2     &  -8.1   \\
Implicit (0+2.125)  & Explicit (3+0)  &  -6.5 & -3.8    &  -6.2  & -10.4  &  -6.7  &  -8.6  & -3.0     &  -6.5   \\
Implicit (0+2.0625) & Explicit (3+0)  &  -4.0 &  2.8    &  -1.9  &  -2.2  &   0.8  &  -2.3  & -2.3     &  -1.3   \\
\midrule
Hybrid (2+4)        & Explicit (3+0)  & -15.1 & -13.2   & -15.4  & -23.8  & -19.9  & -15.8  &  -9.1    & -16.0   \\
Hybrid (2+0.1875)   & Explicit (3+0)  & -13.6 &  -9.6   & -14.8  & -21.4  & -18.0  & -18.4  & -13.1    & -15.6   \\
Hybrid (2+0.125)    & Explicit (3+0)  & -13.0 & -11.2   & -14.1  & -19.1  & -16.3  & -18.1  & -11.0    & -14.7   \\
Hybrid (2+0.0625)   & Explicit (3+0)  &  -9.1 &  -5.4   &  -9.3  & -13.9  & -11.7  & -13.4  &  -6.7    &  -9.9   \\
\midrule
Hybrid (2+0.125)    & Implicit (0+51) & -12.6 & -15.5   & -20.1  & -30.2  & -27.8  & -20.6  & -10.5    & -19.6   \\
Hybrid (2+0.125)    & Implicit (0+6)  & -10.0 & -12.6   & -15.8  & -24.5  & -21.9  & -13.0  &  -8.6    & -15.2   \\
Hybrid (2+0.125)    & Implicit (0+5)  & -9.2  & -13.0   & -14.9  & -24.6  & -21.4  & -14.6  &  -7.4    & -15.0   \\
Hybrid (2+0.125)    & Implicit (0+4)  & -5.9  &  -9.0   & -13.5  & -19.5  & -17.6  &  -1.9  &  -4.3    & -10.2   \\
\midrule
Hybrid (2+0.125)    & Hybrid (3+48)   & -17.3 & -16.3   & -21.0  & -30.3  & -27.3  & -25.7  & -15.1    & -21.9   \\
Hybrid (2+0.125)    & Hybrid (3+3)    & -16.5 & -15.7   & -20.8  & -29.1  & -26.8  & -23.3  & -14.7    & -21.0   \\
Hybrid (2+0.125)    & Hybrid (3+2)    & -17.6 & -15.9   & -20.3  & -29.1  & -25.9  & -26.6  & -14.8    & -21.5   \\
Hybrid (2+0.125)    & Hybrid (3+1)    & -12.7 & -10.4   & -14.8  & -23.6  & -19.3  & -18.9  &  -9.1    & -15.5   \\
\bottomrule
\end{tabular}
\end{table*}
\begin{table*}[h]
\fontsize{9pt}{9pt}\selectfont

\caption{BD-rate (\%) comparison of different buffering strategies with different buffer sizes\new{, using BT.709 for color space conversion.} The anchor employs explicit buffering in both motion and inter-frame coding. \new{The values in parentheses, as in Table~\ref{table:buffer_size}, indicate the number of full-resolution feature maps buffered for coding one input flow map or frame (explicit + implicit).}
}
%
\centering
\label{table:ablation_BT709}
\setlength{\tabcolsep}{5pt}
\begin{tabular}{l|l|rrrrrrr|r}
\toprule
Motion            & Inter         & UVG   & MCL-JCV & HEVC-B & HEVC-C & HEVC-D & HEVC-E & HEVC-RGB & Average \\
\midrule
Explicit (2+0)      & Explicit (3+0)  & 0     & 0       & 0      & 0      & 0      & 0      & 0        & 0       \\
\midrule
Implicit (0+4)      & Explicit (3+0)  & -10.0 & -8.1    & -10.5  & -14.5  & -9.0   & -10.8  & -8.8     & -10.2   \\
Implicit (0+2.1875) & Explicit (3+0)  &  -6.5 & -5.9    &  -8.0  & -10.6  & -6.0   &  -9.4  & -7.2     &  -7.7   \\
Implicit (0+2.125)  & Explicit (3+0)  &  -5.5 & -3.6    &  -5.8  &  -9.3  & -5.1   &  -7.2  & -3.0     &  -5.6   \\
Implicit (0+2.0625) & Explicit (3+0)  &  -3.7 & -0.1    &  -1.6  &  -2.2  &  1.1   &  -1.5  & -2.3     &  -1.5   \\
\midrule
Hybrid (2+4)        & Explicit (3+0)  & -13.3 & -11.5   & -12.8  & -19.1  & -15.0  & -14.4  &  -9.1    & -13.6   \\
Hybrid (2+0.1875)   & Explicit (3+0)  & -12.9 & -11.4   & -13.6  & -18.3  & -14.6  & -17.0  & -13.1    & -14.4   \\
Hybrid (2+0.125)    & Explicit (3+0)  & -12.2 & -10.3   & -12.4  & -15.5  & -12.2  & -18.7  & -11.0    & -13.2   \\
Hybrid (2+0.0625)   & Explicit (3+0)  &  -8.7 &  -6.7   &  -8.3  & -10.7  &  -7.8  & -13.6  &  -6.7    &  -8.9   \\
\midrule
Hybrid (2+0.125)    & Implicit (0+51) & -12.7 & -14.0   & -17.1  & -26.2  & -23.4  & -18.9  & -10.5    & -17.5   \\
Hybrid (2+0.125)    & Implicit (0+6)  & -9.4  & -10.7   & -13.0  & -20.3  & -17.6  & -12.3  & -8.6     & -13.1   \\
Hybrid (2+0.125)    & Implicit (0+5)  & -9.9  & -12.0   & -12.9  & -21.2  & -17.7  & -14.1  & -7.4     & -13.6   \\
Hybrid (2+0.125)    & Implicit (0+4)  & -6.9  &  -7.7   & -11.3  & -16.5  & -14.8  &  -2.6  & -4.3     &  -9.2   \\
\midrule
Hybrid (2+0.125)    & Hybrid (3+48)   & -17.7 & -15.6   & -18.8  & -26.6  & -24.0  & -25.4  & -15.1    & -20.5   \\
Hybrid (2+0.125)    & Hybrid (3+3)    & -16.6 & -14.3   & -18.6  & -25.4  & -23.3  & -23.5  & -14.7    & -19.5   \\
Hybrid (2+0.125)    & Hybrid (3+2)    & -17.5 & -14.6   & -18.3  & -25.3  & -22.8  & -26.8  & -14.8    & -20.0   \\
Hybrid (2+0.125)    & Hybrid (3+1)    & -13.0 &  -9.2   & -13.2  & -19.8  & -16.2  & -19.2  &  -9.1    & -14.2   \\
\bottomrule
\end{tabular}
\end{table*}
\begin{table*}[h]
\fontsize{9pt}{9pt}\selectfont
\caption {BD-rate (\%) comparison of longer sequence training impact on three buffering strategies for inter-frame coding\new{, using BT.709 for color space conversion.} The anchor is the variant employing explicit buffering in both motion and inter-frame coding. \new{The values in parentheses, as in Table~\ref{table:buffer_size}, indicate the number of full-resolution feature maps buffered for coding one input flow map or frame (explicit + implicit).} 
}
%
\centering
\label{table:long_sequence_709}
\setlength{\tabcolsep}{4.5pt}
\begin{tabular}{l|l|r|rrrrrrr|r}
\toprule                   
Motion                            & Inter                         & \# Frame & UVG   & MCL-JCV & HEVC-B & HEVC-C & HEVC-D & HEVC-E & HEVC-RGB & Average \\ 
\midrule
\multirow{2}{*}{Hybrid   (2.125)} & \multirow{2}{*}{Explicit (3)} & 5        & -12.2 & -10.3   & -12.4  & -15.5  & -12.2  & -18.7  & -11.0    & -13.2   \\ 
                                  &                               & 10       & -18.5 & -15.2   & -14.5  & -17.6  & -14.0  & -19.9  & -13.2    & -16.1   \\ 
\midrule
\multirow{2}{*}{Hybrid (2.125)}   & \multirow{2}{*}{Implicit (5)} & 5        & -9.9  & -12.0   & -12.9  & -21.2  & -17.7  & -14.1  & -7.4     & -13.6   \\ 
                                  &                               & 10       & -20.6 & -20.4   & -18.3  & -25.3  & -19.3  & -12.8  & -12.6    & -18.5   \\ 
\midrule
\multirow{2}{*}{Hybrid (2.125)}   & \multirow{2}{*}{Hybrid (2)}   & 5        & -17.5 & -14.6   & -18.3  & -25.3  & -22.8  & -26.8  & -14.8    & -20.0   \\ 
                                  &                               & 10       & -24.5 & -21.7   & -22.5  & -29.5  & -25.8  & -28.8  & -20.3    & -24.7   \\ 
\bottomrule
\end{tabular}
\label{tab:long_sequence}
\end{table*}


\begin{table*}[]
\fontsize{9pt}{9pt}\selectfont
\caption{BD-rate (\%) comparison between our HyTIP and the state-of-the-art methods in terms of PSNR-RGB, using BT.709 for color space conversion. The anchor is VTM 17.0. Negative BD-rates suggest bitrate savings.}
\label{table:sota_psnr_bt709_gop32}
\centering
\begin{tabular}{l|rrrrrrr|r}
\toprule
                        & UVG   & MCL-JCV & HEVC-B & HEVC-C & HEVC-D & HEVC-E & HEVC-RGB & Average \\
\midrule
VTM~\cite{vtm}          & 0.0   & 0.0     & 0.0    & 0.0    & 0.0    & 0.0    & 0.0      & 0.0     \\
HM~\cite{hm}            & 26.0  & 35.7    & 31.9   & 29.9   & 29.8   & 31.9   & 29.6     & 30.7    \\
MaskCRT~\cite{maskcrt}  & 6.4   & 19.1    & 9.2    & 29.0   & 4.7    & 26.1   & -7.2     & 12.5    \\
DCVC-TCM~\cite{tcm}     & 35.0  & 39.3    & 34.1   & 62.5   & 25.5   & 70.4   & 19.2     & 40.9    \\
DCVC-HEM~\cite{hem}     & -6.2  & -0.7    & -0.4   & 19.4   & -5.3   & 9.3    & -14.2    & 0.3     \\
DCVC-DC~\cite{dc}       & -24.0 & -18.4   & -16.4  & -11.6  & -29.0  & -25.4  & -32.1    & -22.4   \\
DCVC-FM~\cite{fm}       & -24.0 & -15.0   & -16.5  & -14.9  & -31.1  & -32.0  & -23.1    & -22.4   \\
HyTIP (Ours)           & -25.8 & -17.0   & -18.8  & -11.0  & -27.9  & -13.9  & -31.3    & -20.8   \\  
\bottomrule
\end{tabular}
\label{tab:SOTA_RD-PSNR_RGB_I32_BT709}
\end{table*}

\begin{table*}[]
\fontsize{9pt}{9pt}\selectfont
\caption{BD-rate (\%) comparison between our HyTIP and the state-of-the-art methods in terms of MS-SSIM-RGB, using BT.709 for color space conversion. The anchor is VTM 17.0. Negative BD-rates suggest bitrate savings.}
\label{table:sota_msssim_bt709_gop32}
\centering
\begin{tabular}{l|rrrrrrr|r}
\toprule
                       & UVG   & MCL-JCV & HEVC-B & HEVC-C & HEVC-D & HEVC-E & HEVC-RGB & Average \\
\midrule
VTM~\cite{vtm}          & 0.0   & 0.0     & 0.0    & 0.0    & 0.0    & 0.0    & 0.0      & 0.0     \\
HM~\cite{hm}            & 21.4  & 31.5    & 28.9   & 29.0   & 29.7   & 29.6   & 26.1     & 28.0    \\
MaskCRT~\cite{maskcrt}  & -23.0 & -30.8   & -38.6  & -29.0  & -43.4  & -31.9  & -43.8    & -34.4   \\
DCVC-TCM~\cite{tcm}     & -9.5  & -22.5   & -26.5  & -17.4  & -33.8  & -18.7  & -27.7    & -22.3   \\
DCVC-HEM~\cite{hem}     & -28.8 & -43.3   & -47.7  & -40.4  & -52.8  & -53.1  & -46.3    & -44.6   \\
DCVC-DC~\cite{dc}       & -36.4 & -50.4   & -55.7  & -51.4  & -61.2  & -66.0  & -56.4    & -53.9   \\
HyTIP (Ours)            & -38.9 & -50.5   & -56.4  & -50.2  & -59.5  & -62.6  & -58.5    & -53.8   \\  
\bottomrule
\end{tabular}
\label{tab:SOTA_RD-MS-SSIM-RGB_I32_BT709}
\end{table*}



\begin{table*}[t!]
\fontsize{9pt}{9pt}\selectfont
\centering
\caption{Training procedure. MENet, TCM, FA (frame type adaptation), CTM, and gain units ($s^{enc}_t, s^{dec}_t, s^{recon}_t, s^{tcm}_t$ in Fig.~\ref{fig:motion}, Fig.~\ref{fig:TCM}, and Fig.~\ref{fig:inter}) represent the motion estimation network, the temporal context mining module in the inter-frame codec, the hierarchical quality structure~\cite{fm}, the channel transform module~\cite{maskcrt}, and the variable-rate modules, respectively. EPA is the error propagation aware training in~\cite{epa}. Ref represents the characteristic of reference temporal information in the inter-frame codec.}

\setlength\tabcolsep{2pt}
\label{tab:training}
\begin{tabular}{lcllcc}

\toprule
Phase  & \# Frames & Training Modules  & Loss  & lr   & Epoch \\
\midrule

\begin{tabular}[l]{@{}l@{}} Motion Coding \\ (Ref: Explicit) \end{tabular}  & 3    &  Motion codec & \begin{tabular}[l]{@{}l@{}} $R^{motion}_t +$ \\ $ \lambda \times D(x_t, warp(x_{t-1}, \hat{f}_t))$ \end{tabular} & 1e-4 & 8    \\\midrule
\begin{tabular}[l]{@{}l@{}} Motion Compensation \\ (Ref: Explicit) \end{tabular}   & 3    & TCM                          & $\lambda \times D(x_t, x_c)$                                  & 1e-4 & 10    \\\midrule
\begin{tabular}[l]{@{}l@{}} Inter-frame Coding \\ (Ref: Explicit) \end{tabular}    & 2    & Inter-frame codec and Mask Generator & $R_t + \lambda \times D(x_t, \hat{x}_t)$                            & 1e-4 & 2   \\\midrule
\begin{tabular}[l]{@{}l@{}} Motion Compensation \\ (Ref: Explicit) \end{tabular}   & 3    & TCM                           & $R_t + \lambda \times \frac{D(x_t, x_c) + D(x_t, \hat{x}_t)}{2} $ & 1e-4 & 3    \\\midrule
\multicolumn{1}{l}{\multirow{2}{*}{\begin{tabular}[l]{@{}l@{}} Inter-frame Coding \\ (Ref: Explicit) \end{tabular} }} & 3    & \multicolumn{1}{l}{\multirow{2}{*}{\begin{tabular}[l]{@{}l@{}} All modules except MENet, motion codec, \\ and 3x3 Conv in Fig.~\ref{fig:overview} \end{tabular}}} & $R_t + \lambda \times D(x_t, \hat{x}_t)$    & 1e-4 & 8     \\
\multicolumn{1}{l}{}                                                    & 5    & \multicolumn{1}{l}{}                                                                              & $R_t + \lambda \times D(x_t, \hat{x}_t)$   & 1e-4 & 5     \\\midrule
\multicolumn{1}{l}{\multirow{2}{*}{\begin{tabular}[l]{@{}l@{}} Fine-tuning \\ (Ref: Explicit) \end{tabular} }}         & 3    & \multicolumn{1}{l}{\multirow{2}{*}{\begin{tabular}[l]{@{}l@{}} All modules except MENet and 3x3 Conv \\ in Fig.~\ref{fig:overview}  \end{tabular}}}  & $R_t + \lambda \times D(x_t, \hat{x}_t)$  & 1e-4 & 6    \\
\multicolumn{1}{l}{}                                                    & 5    & \multicolumn{1}{l}{}                                                                 & $R_t + \lambda \times D(x_t, \hat{x}_t)$  & 1e-4 & 5    \\\midrule
\begin{tabular}[l]{@{}l@{}} Feature Generation \\ (Ref: Hybrid) \end{tabular} & 3    & 3x3 Conv in Fig.~\ref{fig:overview}                                 & $R_t + \lambda \times D(x_t, \hat{x}_t)$                            & 1e-4 & 3    \\\midrule
\begin{tabular}[l]{@{}l@{}} Motion Compensation \\ (Ref: Hybrid) \end{tabular} & 3    & TCM                           & $R_t + \lambda \times \frac{D(x_t, x_c) + D(x_t, \hat{x}_t)}{2} $ & 1e-4 & 4    \\\midrule
\multicolumn{1}{l}{\multirow{2}{*}{\begin{tabular}[l]{@{}l@{}} Inter-frame Coding \\ (Ref: Hybrid) \end{tabular} }}   & 3    & \multicolumn{1}{l}{\multirow{2}{*}{All modules except MENet and motion codec}}      & $R_t + \lambda \times D(x_t, \hat{x}_t)$    & 1e-4 & 8   \\
\multicolumn{1}{l}{}                                                    & 5    & \multicolumn{1}{l}{}                                                                & $R_t + \lambda \times D(x_t, \hat{x}_t)$    & 1e-4 & 4    \\\midrule
\multicolumn{1}{l}{\multirow{2}{*}{\begin{tabular}[l]{@{}l@{}} Fine-tuning \\ (Ref: Hybrid) \end{tabular} }}           & 3    & \multicolumn{1}{l}{\multirow{2}{*}{All modules except MENet}}    & $R_t + \lambda \times D(x_t, \hat{x}_t)$                            & 1e-4 & 3    \\
\multicolumn{1}{l}{}                                                    & 5    & \multicolumn{1}{l}{}      & $R_t + \lambda \times D(x_t, \hat{x}_t)$    & 1e-4 & 4    \\\midrule
\multicolumn{1}{l}{\multirow{2}{*}{\begin{tabular}[l]{@{}l@{}} Fine-tuning with EPA \\ (Ref: Hybrid) \end{tabular} }}  & 5    & All modules except MENet     & $R_t + \lambda \times D(x_t, \hat{x}_t)$                            & 1e-5 & 4    \\
\multicolumn{1}{l}{}                                                    & 5    & All modules                               & $R_t + \lambda \times D(x_t, \hat{x}_t)$                            & 1e-5 & 4    \\
\midrule
\multicolumn{1}{l}{\multirow{2}{*}{\begin{tabular}[l]{@{}l@{}} FA (Ref: Hybrid) \\ FA with EPA (Ref: Hybrid) \end{tabular} }}   & 5    & \multicolumn{1}{l}{\multirow{2}{*}{\begin{tabular}[l]{@{}l@{}} Frame Type Conv layers in Fig.~\ref{fig:TCM}  \end{tabular}}}      & $R_t + \lambda \times D(x_t, \hat{x}_t)$    & 1e-4 & 1   \\
\multicolumn{1}{l}{}                                                    & 5    & \multicolumn{1}{l}{}                                                                & $R_t + \lambda \times D(x_t, \hat{x}_t)$    & 1e-5 & 1    \\\midrule
\multicolumn{1}{l}{\multirow{2}{*}{\begin{tabular}[l]{@{}l@{}} Fine-tuning with EPA \\ (Ref: Hybrid) \end{tabular} }}  & 5    & All modules except MENet     & $R_t + \lambda \times D(x_t, \hat{x}_t)$                            & 1e-5 & 1    \\
\multicolumn{1}{l}{}                                                    & 5    & All modules                               & $R_t + \lambda \times D(x_t, \hat{x}_t)$                            & 1e-5 & 5    \\
\midrule
\begin{tabular}[l]{@{}l@{}} CTM \\ (Ref: Hybrid) \end{tabular} & 5    & CTM in Fig.~\ref{fig:inter}                          & $R_t + \lambda \times D(x_t, \hat{x}_t) $ & 1e-4 & 1    \\
\midrule
\multicolumn{1}{l}{\multirow{2}{*}{\begin{tabular}[l]{@{}l@{}} Fine-tuning with EPA \\ (Ref: Hybrid) \end{tabular} }}  & 5    & All modules except MENet     & $R_t + \lambda \times D(x_t, \hat{x}_t)$                            & 1e-5 & 2    \\
\multicolumn{1}{l}{}                                                    & 5    & All modules                               & $R_t + \lambda \times D(x_t, \hat{x}_t)$                            & 1e-5 & 7    \\
\midrule
\multicolumn{1}{l}{\multirow{2}{*}{\begin{tabular}[l]{@{}l@{}} Context Model Training \\ with EPA (Ref: Hybrid) \end{tabular} }}   & 5    & \multicolumn{1}{l}{\multirow{2}{*}{\begin{tabular}[l]{@{}l@{}} Context Model in Fig.~\ref{fig:inter}\\ Context Model, hyperprior, and decoder in Fig.~\ref{fig:inter}  \end{tabular}}}      & $R_t + \lambda \times D(x_t, \hat{x}_t)$    & 1e-4 & 1   \\
\multicolumn{1}{l}{}                                                    & 5    & \multicolumn{1}{l}{}                                                                & $R_t + \lambda \times D(x_t, \hat{x}_t)$    & 1e-4 & 2    \\\midrule
\multicolumn{1}{l}{\multirow{2}{*}{\begin{tabular}[l]{@{}l@{}} Inter-frame Coding \\ with EPA (Ref: Hybrid) \end{tabular} }} & 5    & Inter-frame codec and Mask Generator & $R_t + \lambda \times D(x_t, \hat{x}_t)$                            & 1e-4 & 3   \\
 & 5    & All modules except MENet and motion codec & $R_t + \lambda \times D(x_t, \hat{x}_t)$                            & 1e-4 & 2   \\\midrule
\multicolumn{1}{l}{\multirow{2}{*}{\begin{tabular}[l]{@{}l@{}} Fine-tuning with EPA \\ (Ref: Hybrid) \end{tabular} }}  & 5    & All modules except MENet     & $R_t + \lambda \times D(x_t, \hat{x}_t)$                            & 1e-5 & 3    \\
\multicolumn{1}{l}{}                                                    & 5    & All modules                               & $R_t + \lambda \times D(x_t, \hat{x}_t)$                            & 1e-5 & 6    \\\midrule

\multicolumn{1}{l}{\multirow{2}{*}{\begin{tabular}[l]{@{}l@{}} Variable-rate Training \\ with EPA (Ref: Hybrid) \end{tabular} }}  & 5    & Gain units in Fig.~\ref{fig:motion}, Fig.~\ref{fig:TCM}, and Fig.~\ref{fig:inter} & $R_t + \lambda \times D(x_t, \hat{x}_t)$                            & 1e-5 & 8    \\
\multicolumn{1}{l}{}                                                    & 5    & All modules                               & $R_t + \lambda \times D(x_t, \hat{x}_t)$                            & 1e-5 & 20    \\\midrule


\multirow{2}{*}{\begin{tabular}[l]{@{}l@{}} Long-sequence Training \\ with EPA (Ref: Hybrid) \end{tabular} } & 7    & All modules  & $R_t + \lambda \times D(x_t, \hat{x}_t)$ & 1e-5 & 1    \\
             & 10   & All modules  & $R_t + \lambda \times D(x_t, \hat{x}_t)$ & 1e-6 & 50   \\
\bottomrule
\end{tabular}
\end{table*}



\section{Configurations of HM 16.25 and VTM 17.0}
\label{sec:configuration}
Following the recommendation from \cite{dc}, we encode videos in YUV444 format. We use the \textit{encoder\_lowdelay\_vtm.cfg} of VTM~\cite{vtm} with the following parameters:

--c \{config file name\}

--InputFile=\{input file name\}

--InputBitDepth=8

--InputChromaFormat=444

--ChromaFormatIDC=444

--InternalBitDepth=10

--OutputBitDepth=8

--DecodingRefreshType=2

--FrameRate=\{frame rate\}

--FrameSkip=0

--SourceWidth=\{width\}

--SourceHeight=\{height\}

--FramesToBeEncoded=96

--Level=4.1

--IntraPeriod=32

--QP=\{qp\}

--BitstreamFile=\{bitstream file name\}

--ReconFile=\{reconstruction file name\}

\vspace{0.2cm}

\noindent Similarly, We use the \textit{encoder\_lowdelay\_main\_rext.cfg} of HM~\cite{vtm} with the following parameters:

--c \{config file name\}

--InputFile=\{input file name\}

--InputBitDepth=8

--InputChromaFormat=444

--ChromaFormatIDC=444

--InternalBitDepth=10

--InternalBitDepthC=10

--OutputBitDepth=8

--OutputBitDepthC=8

--FrameRate=\{frame rate\}

--FrameSkip=0

--SourceWidth=\{width\}

--SourceHeight=\{height\}

--FramesToBeEncoded=96

--Level=4.1

--IntraPeriod=32

--QP=\{qp\}

--BitstreamFile=\{bitstream file name\}

--ReconFile=\{reconstruction file name\}



\end{document}